\documentclass[
reprint,
superscriptaddress,
amsmath,amssymb,
aps,
prb,
floatfix,
]{revtex4-1}

\usepackage{xcolor}
\usepackage{hyperref}
\usepackage{graphicx}
\usepackage{braket}
\usepackage[normalem]{ulem}
\usepackage{gensymb}
\usepackage{mathtools}
\usepackage{nccmath}
\usepackage{bbold}
\usepackage{relsize}
\usepackage{amsmath}
\usepackage{mathrsfs}
\usepackage{physics}

\definecolor{webblue}{HTML}{000099}
\hypersetup{
colorlinks=true,linkcolor=webblue,citecolor=webblue,
filecolor=webblue,urlcolor=webblue,
pdftitle={Magnetic Order in Moire Graphene},
}

\newcommand{\vet}[1]{\mathbf{#1}}
\newcommand{\sublabel}[2]{(#1)\hfill{}#2\hfill{}\phantom{(#1)}\\}

\usepackage{makerobust}
\newcommand{\makeauthor}[2]{\newcommand{#1}[1]{{%
  \protect%
  \color{#2}{%
    \bfseries\begingroup\escapechar=-1\edef\x{\endgroup\string#1}\x:%
  }\itshape{} ##1}}%
  \MakeRobustCommand#1}
\makeauthor{\lk}{purple}
\makeauthor{\af}{blue}
\begin{document}

\title{Magnetic Ordering in Moir\'e Graphene Multilayers from a Continuum Hartree+U Approach}

\author{Christopher T. S. Cheung}
\affiliation{Departments of Physics and Materials and the Thomas Young center for Theory and Simulation of Materials, Imperial College London, South Kensington Campus, London SW7 2AZ, UK\\}
\author{Valerio Vitale}
\affiliation{Departments of Physics and Materials and the Thomas Young center for Theory and Simulation of Materials, Imperial College London, South Kensington Campus, London SW7 2AZ, UK\\}
\affiliation{Dipartimento di Fisica, Universit\`a degli Studi di Trieste, strada costiera 11, 34151 Trieste, Italy}
\affiliation{CNR-IOM--Istituto Officina dei Materiali, National Research Council of Italy, c/o SISSA Via Bonomea 265, Trieste IT-34136, Italy}
\author{Lennart Klebl}
\affiliation{Institut f\"ur Theoretische Physik und Astrophysik and W\"urzburg-Dresden Cluster of Excellence ct.qmat, Universit\"at W\"urzburg, 97074 W\"urzburg, Germany}
\author{Ammon Fischer}
\affiliation{Max Planck Institute for the Structure and Dynamics of Matter, Center for Free Electron Laser Science, 22761 Hamburg, Germany}
\affiliation{Institute for Theory of Statistical Physics, RWTH Aachen University, and JARA Fundamentals of Future Information Technology, 52062 Aachen, Germany}
\author{Dante M. Kennes}
\affiliation{Max Planck Institute for the Structure and Dynamics of Matter, Center for Free Electron Laser Science, 22761 Hamburg, Germany}
\affiliation{Institute for Theory of Statistical Physics, RWTH Aachen University, and JARA Fundamentals of Future Information Technology, 52062 Aachen, Germany}
\author{Arash A. Mostofi}
\affiliation{Departments of Physics and Materials and the Thomas Young center for Theory and Simulation of Materials, Imperial College London, South Kensington Campus, London SW7 2AZ, UK\\}
\author{Johannes Lischner}
\affiliation{Departments of Physics and Materials and the Thomas Young center for Theory and Simulation of Materials, Imperial College London, South Kensington Campus, London SW7 2AZ, UK\\}
\author{Zachary A. H. Goodwin}
\affiliation{Departments of Physics and Materials and the Thomas Young center for Theory and Simulation of Materials, Imperial College London, South Kensington Campus, London SW7 2AZ, UK\\}
\affiliation{John A. Paulson School of Engineering and  Applied Sciences, Harvard  University, Cambridge, MA 02138, USA}
\affiliation{Department of Materials, University of Oxford, Parks Road, Oxford OX1 3PH, United Kingdom}

\date{\today}

\begin{abstract}
Recently, symmetry-broken ground states, such as correlated insulating states, magnetic order and superconductivity, have been discovered in twisted bilayer graphene (tBLG) and twisted trilayer graphene (tTLG) near the so-called magic-angle. Understanding the magnetic order in these systems is challenging\textcolor{black}{, however}, as atomistic methods become extremely expensive near the magic angle and continuum approaches fail to capture important atomistic details. In this work, we develop \textcolor{black}{an approach to incorporate short-ranged Hubbard interactions self-consistently in a continuum model.} \textcolor{black}{In addition, we include} long-ranged Coulomb interactions~\textcolor{black}{, which are known to be important when doping the flat bands of tBLG and tTLG }. Therefore, \textcolor{black}{for the first time,} magnetic order in moir\'e graphene multilayers \textcolor{black}{is self-consistently explored in a continuum model with atomistic detail}. With this approach, we perform a systematic analysis of the magnetic phase diagram of tBLG as a function of doping level and twist angle, near the magic angle. \textcolor{black}{Our} results are consistent with previous perturbative atomistic Hartree+U calculations. Furthermore, we \textcolor{black}{investigated} magnetic order \textcolor{black}{of} tTLG, \textcolor{black}{which were found to be} similar to those in tBLG. In the future, the developed \textcolor{black}{continuum model} can be utilized to investigate magnetic ordering tendencies from short-range exchange interactions in other moir\'e graphene multilayers as a function of doping, twist angle, screening environment, among other variables.
\end{abstract}

\maketitle

\section{Introduction}
\label{sec:intro}

Since the discovery of correlated insulating states~\cite{Cao2018_corr_insul} and superconductivity~\cite{Cao2018_supercond} in magic-angle twisted bilayer graphene (tBLG), moir\'e materials have emerged as an excellent platform for understanding flat-band physics and strongly-correlated electrons. This is demonstrated by the unprecedented variety of exotic electronic states that have been experimentally observed in tBLG, such as global charge-ordered stripe phases~\cite{Jiang2019}, ferromagnetic chiral edge states~\cite{Sharpe2019}, orbital magnetism \cite{Lu2019}, among other experimental and theoretical findings \cite{Cao2016,Carr2017,Zou2018,Choi2019,Lu2019,Jiang2019,Kang2019,Kerelsky2019,Lucignano2019,Polshyn2019,Xie2019,Yankowitz2019,Yoo2019,Andrei2020,Cao2020,Carr2020,Balents2020,Bultinck2020,Nuckolls2020,Xie2020,Zhang2020,Zondiner2020,Cao2021,Choi2021,Das2021,Liu2021,Kennes2021,Rozen2021,Saito2021,Oh2021,Wu2021,Xie2021}. After these discoveries, flat-band physics and strong correlations have been discovered in moir\'e materials other than tBLG. For example, still using graphene as a building block, twisted trilayer graphene~\cite{Hao21,Fischer_unconv_supercond_2022} (tTLG), twisted double bilayer graphene (tDBLG)~\cite{Crommie2021,Carmen2021,He2021,BIBI}, among other systems~\cite{PHD_7}\textcolor{black}{,} have been actively investigated, \textcolor{black}{where} correlated insulating states and superconductivity have \textcolor{black}{also} been found. \textcolor{black}{Moreover, this prompted} multilayer graphene systems without a twist \textcolor{black}{to be} revisited\textcolor{black}{, leading, for instance, to the discovery of} superconductivity and magnetism in rhombohedral graphene multilayers~\cite{seiler2022quantum,zhou2021superconductivity, zhou2022isospin,zhang2023enhanced,tsui2024direct,han2024sig,lu2024fractional}. In addition to twisted graphitic materials, twisted bilayers and multilayers consisting of other 2D materials have also been investigated, such as semiconducting transition metal dichalcogenides (TMDs)~\cite{Vitale22,UFSS,TITMD,xia2025superconductivity,guo2025superconductivity,park2023observation}, metallic TMDs~\cite{GoodwinMM22,Cheung24}, hexagonal boron nitride~\cite{hBN_Xian2019,hBN_Walet2021}, among other examples~\cite{Kennes2021}. 

In tBLG, flat bands near the Fermi energy \textcolor{black}{were} predicted by \textcolor{black}{Bistrizer and MacDonald's} continuum model \textcolor{black}{at specific twist angles, referred to as magic angles}~\cite{BMD_tBLG_cont_2011}. \textcolor{black}{Since the discovery of broken symmetry states at the magic-angle,} theoretical effort has \textcolor{black}{largely focused on} incorporating interactions into the electronic Hamiltonian of this continuum model\textcolor{black}{, in an attempt to understand these phases}. Notably, Guinea \textit{et al.} incorporated long-ranged Hartree interactions into the continuum model, predicting strong distortions of the flat bands near the magic angle \cite{Guinea_Hart_distort_2018,Cea_Hart_Pin_2019,Cea_Hart_Dist_2020,Cea_Hart_revw_2022}. Long-ranged exchange, i.e., Fock, interactions have also been included in the continuum model, giving rise to various broken symmetry states in reasonable agreement with experimental measurements~\cite{SCHFC,Cea_Hart_Dist_2020,Cea_Hart_revw_2022}. Alternative approaches based on the Wannier functions of the flat bands have also had success in explaining experimental measurements~\cite{SMLWF,Kang2019,MLWO,PHD_3,PHD_1,PHD_2}. However, neither of these techniques has clearly determined the role of short-ranged interactions, as separating the effects of short-range and long-range interactions in these methods is difficult.

Atomistic models, on the other hand, can naturally include short-ranged interactions, such as the onsite repulsion between electrons in p$_z$ orbitals~\cite{GonzalezArraga2017,Bezanilla2019,Stauber2021,Vahedi2021}. The magnetic phases of tBLG have been studied using perturbative random phase approximation (RPA) atomistic calculations at the magic angle by Klebl~\textit{et al.}~\cite{Klebl_mag_order_tBLG_2019,Goodwin_mag_tBLG_2021,fischer2021spin}, which predicted the existence of emergent moir\'e-scale modulated antiferromagentic and ferromagnetic states~\cite{Klebl_mag_order_tBLG_2019}. Klebl~\textit{et al.}~\cite{Klebl_mag_order_tBLG_2019,Goodwin_mag_tBLG_2021} have also mapped the magnetic phase diagram of tBLG over a larger range of twist angles at \textcolor{black}{integer} doping levels up to $\pm3$ electrons per moir\'e unit cell, taking into account long-ranged Hartree interactions (in a perturbative way)~\cite{Goodwin_mag_tBLG_2021}. However, a major challenge presented by atomistic tight-binding approaches is the computational cost of reaching converged solutions for small twist-angle systems~\cite{GonzalezArraga2017,Bezanilla2019,Stauber2021,Vahedi2021}. Motivated by the low computational cost of continuum models, and the success of atomistic models in capturing short-ranged interactions, Jimeno-Pozo~\textit{et al.}~\cite{Alejandro23} developed a hybrid approach to incorporate \textcolor{black}{atomistic onsite p$_z$} Hubbard interactions into the continuum model ~\cite{Klebl_mag_order_tBLG_2019,Goodwin_mag_tBLG_2021}. The magnetic orders predicted by the atomistic models were incorporated into the continuum model through a sublattice-polarized moir\'e expansion, similar to how Hartree interactions are included in the continuum model~\cite{Alejandro23}. Those calculations, however, were only perturbative. 


In this work, we develop and perform self-consistent Hartree+U calculations in the continuum model of moir\'e graphene multilayers. Using this methodology, we compute the self-consistent order parameters of \textcolor{black}{several} magnetic instabilities over a range of doping levels in the flat bands and a range of twist angles near the magic angle. \textcolor{black}{These calculations provide} insight into the regions of stability for each competing magnetic ground state in tBLG. The \textcolor{black}{developed} approach is general, which we demonstrate by predicting quasiparticle band structures for twisted symmetric trilayer graphene (tTLG), with different possible magnetic orders, where we found similarities to tBLG. We hope this method will open up avenues \textcolor{black}{to} include both short and long range exchange interactions, to determine their relative importance. Moreover, the developed method could also be used for other moir\'e graphene multilayers, or moir\'e systems comprising \textcolor{black}{of} other 2D materials. 

\section{Methods}\label{sec:methods}

We take a combined atomistic-continuum \textcolor{black}{approach here}. First, we determine the leading magnetic instabilities using an RPA theory based on the atomistic tight-binding model. As these magnetic instabilities have already been discussed in depth elsewhere, we refer the reader to Refs.~\cite{Klebl_mag_order_tBLG_2019,Goodwin_mag_tBLG_2021} for the details of these calculations. These leading magnetic instabilities are then included in the continuum model by approximating their form with a sum of trigonometric functions. The reader should refer to Ref.~\cite{Alejandro23} for a more detailed description of this procedure. In the remainder of the Methods Section, we outline the self-consistent continuum model approach developed here, which builds on the previously mentioned work.

\subsection{Continuum model for tBLG}
\label{app:cont_tBLG}

The atomic structure of tBLG is obtained by rotating the top and bottom layers by $\pm \theta/2$, where $\theta$ is the twist angle. As we build on atomistic methods, we choose twist angles \textcolor{black}{with commensurate moir\'e unit cells}, described by the moir\'e lattice vector $\vet{R}_1=m\vet{a}_1+n\vet{a}_2$, with $(m,n)$ being non-equal integers, and $\vet{a}_1=(\sqrt{3}/2,-1/2)a_0$, $\vet{a}_2=(\sqrt{3}/2,1/2)a_0$ are the graphene unit cell vectors. The lattice constant $a_0$ is taken as 2.46~$\mathrm{\AA}$.

The low-energy electronic properties of tBLG, without any electron-electron interactions, are described with a Bistritzer-MacDonald continuum model~\cite{BMD_tBLG_cont_2011} 
\begin{equation}
    \mathcal{H}_0^{\chi}(\mathbf{k};\theta)=
    \begin{pmatrix}
        \mathcal{H}^{1,\chi}(\mathbf{k};\theta) & \mathcal{U}^{\chi,\dagger} \\
        \mathcal{U}^{\chi} & \mathcal{H}^{2,\chi}(\mathbf{k};\theta) \\ 
    \end{pmatrix},
\label{eq:H_cont_tBLG}
\end{equation}
where $\mathcal{H}^{l,\chi}(\mathbf{k};\theta)$ is the $l$-th intralayer Dirac Hamiltonian for valley $\chi$, \textcolor{black}{where} $\chi = 1(-1)$ for the $K$($K'$)-valley \textcolor{black}{, and $\textbf{k}$ is the moir\'e-scale crystal momentum.} \textcolor{black}{In the continuum model, the Bloch eigenfunctions are expanded in terms of moir\'e plane waves}. Hence, the $(N,M)$-th \textcolor{black}{block} of the \textcolor{black}{$l$-th intralayer Dirac Hamiltonian} is a $2\times2$ submatrix describing the sublattice coupling as 
\begin{equation}
    \mathcal{H}^{l,\chi}_{N,M}(\mathbf{k};\theta) = 
    \begin{pmatrix}
        0 & \mathnormal{h}^{l,\chi}(\mathbf{k}+\mathbf{G}_{N,M};\theta) \\
        h^{l,\chi}(\mathbf{k}+\mathbf{G}_{N,M};\theta) & 0 
    \end{pmatrix},
\label{eq:H_lay}
\end{equation}
where $\mathbf{G}_{N,M}$ is a moir\'e reciprocal lattice vector given by $\mathbf{G}_{N,M} = N\mathbf{G}_1+M\mathbf{G}_2$, with $\mathbf{G}_1 = 2\pi/L(1/\sqrt{3},1)$ and $\mathbf{G}_2 = 4\pi/L(-1/\sqrt{3},0)$, where $L = a_0/(2\sin(\theta/2)$ is the length of the moir\'e cell vectors. The integers $N,M$ run over the set $\mathcal M$ that indexes moir\'e reciprocal lattice vectors such that they satisfy
\begin{equation}
    \mathcal M = \{ (N,M)\in\mathbb{Z}^2 ~|~ \norm{\mathbf{G}_{N,M}} \leq p_{\mathrm{max}}\norm{\mathbf{G}_1} \} \,.
\end{equation}
Here $p_{\mathrm{max}}$ is an integer controlling the cutoff radius in reciprocal space, and was taken as 5. 

The functions $h^{l,\chi}(\mathbf{q};\theta)$ are given by 
\begin{equation}
    h^{l,\chi}(\vet{q};\theta) = \chi\hbar v_F (\vet{q}-\chi \vet{K}_l)\cdot\boldsymbol{\tau}_{\theta}^{l,\chi},
    \label{eq:H_dirac}
\end{equation}
where $\vet{K}_l$ is the Dirac point associated with the $l$-th layer, $v_F=\sqrt{3}ta_0/2\hbar$ is the Fermi velocity in graphene, with $t$ the hopping amplitude between localized $p_z$ orbitals, and 
\begin{equation}
    \boldsymbol{\tau}_{\theta}^{l,\chi} = \exp(i\chi\tau_z\theta_l/2)
    (\tau_x,\chi\tau_y)\exp(-i\chi\tau_z\theta_l/2),
\label{eq:H_rot}
\end{equation}
with $\tau_i$ being the corresponding Pauli matrices, and $\theta_{1,2}=\mp \theta/2$. From now on, we will drop the dependence on $\theta$ in the Dirac Hamiltonians, which will be implicitly assumed.

Moving on to the interlayer coupling Hamiltonians, $\mathcal{U}$~\cite{Cea_Hart_Pin_2019,Cea_Hart_Dist_2020,Cea_Hart_revw_2022}, we have the $(N,M)$-th submatrix given by
\begin{equation}
\begin{aligned}
    \mathcal{U}^{\chi}_{N,M} = & U_1^{\chi}\delta^{(2)}(\mathbf{G}_{N}-\mathbf{G}_{M}) +\\ 
    & U_2^{\chi}\delta^{(2)}(\mathbf{G}_{N}-\mathbf{G}_{M}-\chi\mathbf{G}_{1})+\\
    & U_3^{\chi}\delta^{(2)}(\mathbf{G}_{N}-\mathbf{G}_{M}-\chi(\mathbf{G}_{1}+\mathbf{G}_{2})),
\end{aligned}
\label{eq:U_KM}
\end{equation}
where $\delta^{(2)}$ is a Kronecker delta function acting in the 2D reciprocal space, with 
\begin{equation}
    \begin{aligned}
    U_1^{\chi} & = 
    \begin{pmatrix}
    u_1 & u_2 \\ u_2 & u_1     
    \end{pmatrix} \\
    U_2^{\chi} & = 
    \begin{pmatrix}
    u_1 & u_2e^{-i\frac{2\pi\chi}{3}} \\ u_2e^{i\frac{2\pi\chi}{3}} & u_1 
    \end{pmatrix} \\
    U_3^{\chi} & = 
    \begin{pmatrix}
    u_1 & u_2e^{i\frac{2\pi\chi}{3}} \\ u_2e^{-i\frac{2\pi\chi}{3}} & u_1 
    \end{pmatrix},
    \end{aligned}
\label{eq:U_123}
\end{equation}
acting on the sublattice subspace. The interlayer tunnelings $u_{1,2}$ are taken as \mbox{$u_1=79.7$~meV} and \mbox{$u_2=97.5$~meV~\cite{MLWO,Cea2020}}.  

For the Hartree electron-electron interactions, we adopt the description proposed by Guinea~\textit{et al.}~\cite{Guinea_Hart_distort_2018}, where the Hartree Hamiltonian submatrix is given by
\begin{equation}
    \mathcal{H}^{H} = \frac{\delta\rho_{G}}{\epsilon L}\mathbb{1}.
    \label{eq:H_hart}
\end{equation}
Here $\epsilon=24$ is the effective dielectric constant describing the external and internal screening~\cite{Cea2019}, \textcolor{black}{and $\delta\rho_{G}$ is the moir\'e scale electron density, referenced to charge neutrality, which needs to be self-consistently determined. Note the Hartree interactions treat tBLG to act as 2D sheet, since $\exp(-d|\textbf{G}|) \approx 1$, where $d \approx 3.35$\AA~ is the interlayer separation, which means this term does not explicitly depend on layer.}

Following Ref.~\cite{Alejandro23}, the mean-field Hubbard interactions between electrons due to magnetic ordering are captured by the inter-sublattice coupling Hamiltonian
\begin{equation}
\begin{aligned}
    \mathcal{H}^{U,l}_{N,M} = & \delta_0^l \mathcal{S} \delta^{(2)}(\mathbf{G}_{N}-\mathbf{G}_{M}) +\\ 
    & \delta_1^l \mathcal{S}\sum_{\mathbf{G}_\star}\delta^{(2)}(\mathbf{G}_{N}-\mathbf{G}_{M}-\mathbf{G}_{\star}),
    \label{eq:H_Hub}
\end{aligned}
\end{equation}
where $\mathbf{G}_\star$ are the first star of reciprocal moir\'e lattice vectors, $\mathcal{S}$ is a $2\times2$ sublattice-coupling matrix describing the magnetic order\textcolor{black}{, the definitions of which are given in Results Section~\ref{sec:results}, and the }magnetic order parameters, $\delta_0^l$ and $\delta_1^l$, are determined through
\begin{equation}
    \delta_0^l = 2U(n^l_{0,\uparrow}-n^l_{0,\downarrow})
    \label{eq:delta0}
\end{equation}
and
\begin{equation}
    \delta_1^l=2U(n^l_{G,\uparrow}-n^l_{G,\downarrow}),
    \label{eq:delta1}
\end{equation}
where $U$ is the Hubbard parameter, the factor of 2 accounts for the valley degeneracy, $n_{0,\sigma}^l$ is the constant contribution to the spin-polarized electron density, \textcolor{black}{and $n_{G,\sigma}^l$ is the moir\'e scale contribution to the spin-polarized electron density. Here we have retained the explicit layer index to be clear that these terms act within a layer.} 

Incorporating the \textcolor{black}{non-interacting Hamiltonian with the} Hartree, $\mathcal{H}^{H}$, and Hubbard, $\mathcal{H}^{U} $, contributions, we obtain the valley- and spin-projected continuum Hamiltonian as
\begin{equation}
\mathcal{H}^{\chi,\sigma} = \mathcal H^H + (-1)^\sigma \mathcal H^U + \mathcal{H}_0^{\chi},
    \label{eq:H_cont_fin}
\end{equation}
where $\sigma={0,1}$ for spin up and down electrons, respectively, and the first two terms are $\propto \mathbb{1}$ in layer space. With the Hamiltonian in~ Eq.\eqref{eq:H_cont_fin}, we solve the eigenvalue equation 
\begin{equation}
    \mathcal{H}^{\chi,\sigma}(\mathbf{k})\mathbf{c}_{\chi,\sigma}(\mathbf{k}) = E^{\chi}_{\sigma}(\mathbf{k})\mathbf{c}_{\chi,\sigma}(\mathbf{k}),
    \label{eq:EigEq_cont}
\end{equation}
where $\mathbf{c}_{\chi,\sigma}(\mathbf{k})$ is the eigenvector and $E^{\chi}_{\sigma}(\mathbf{k})$ are the energy eigenvalues for a given spin $\sigma$.

The order parameter $\delta\rho_{G}$ is computed from the eigenstates \textcolor{black}{through} 
\begin{equation}
\begin{aligned}
      \delta\rho_G = \frac{2}{N_k} & \sum_{\substack{\sigma,\\ n_\mathrm{occ}^{CN}}}  \sum_{\substack{\mathbf{k},\mathbf{G'},\\ \mathbf{k}+\mathbf{G'} = \mathbf{q}}} c_{n,\chi,\sigma}(\mathbf{q})c_{n,\chi,\sigma}^{*}(\mathbf{q}+\mathbf{G}),
    \label{eq:nG}  
\end{aligned}
\end{equation}
where $N_k$ is the number of $k$-points sampled in the mini-Brillouin zone, $n_{\mathrm{occ}}^{CN}$ denotes \textcolor{black}{the sum of occupied states relative to charge neutrality~\cite{EE,Cea2019,Cea2020,Pierre2020}, with $n$ being used as short-hard in the eigenstate coefficients. Note to include temperature effects, the summation over occupied states is replaced with a summation of the flat bands, still relative to the charge neutrality state, and a Fermi occupation function is included~\cite{EE,Cea2019,Cea2020,Pierre2020}. We refrain from further investigating the role of temperature here for simplicity, and focus on the $T=0$~K limit.} 

The \textcolor{black}{moir\'e variation in the} spin-polarized electron density $n_{G,\sigma}$ is \textcolor{black}{calculated} as
\begin{equation}
\begin{aligned}  
n_{G,\sigma}^l = \frac{2}{N_k}\sum_{n_\mathrm{occ}}\sum_{\substack{\mathbf{k},\mathbf{G'},\\ \mathbf{k}+\mathbf{G'} = \mathbf{q}}}c_{n,\sigma}^l(\mathbf{q})c_{n,\sigma}^{*,l}(\mathbf{q}+\mathbf{G}),
\label{eq:nGpspin}
\end{aligned}
\end{equation}
\textcolor{black}{and the atomic unit-cell constant} spin-polarised density is defined as 
\begin{equation}
\begin{aligned}
n_{0,\sigma}^l  = 
\frac{2}{N_k}\sum_{n_\mathrm{occ}}\sum_{\substack{\mathbf{k},\mathbf{G'},\\ \mathbf{k}+\mathbf{G'} = \mathbf{q}}}c^{l}_{n,\sigma}(\mathbf{q}) c_{n,\sigma}^{*,l}(\mathbf{q}).
\label{eq:n0spin}
\end{aligned}
\end{equation}
Note that in calculating these spin densities, all bands below the Fermi level are taken into account, in contrast to the expression for $\delta\rho_G$ in Eq.~\eqref{eq:nG} \textcolor{black}{where it is relative to charge neutrality in the flat bands}, and the factor of 2 in these equations comes from a valley degeneracy. \textcolor{black}{Again, to introduce finite temperature effects, the summation over occupied bands is replaced with a summation over all bands and a Fermi function is introduced.}

The order parameters $\delta_0$, $\delta_1$, and $\delta\rho_G$ are obtained self-consistently. The convergence criterion was chosen so that the total difference between the successive values of all 3 order parameters is less than $10^{-6}$. We note that occasionally the criterion is relaxed for metallic systems. Nonetheless, the \textcolor{black}{order} parameters are \textcolor{black}{all} converged to at least 3 significant figures. 

\subsection{Continuum Model for tTLG}
\label{app:cont_ttlg}

In the case of tTLG, the external layers are twisted symmetrically with respect to the central layer, resulting in a mirror-symmetric system. The lattice vectors and commensurate integers which describe the system remain the same as for tBLG. The spin-valley-projected Hamiltonian for tTLG is similar to that for \textcolor{black}{tBLG,} with a number of \textcolor{black}{slight} modifications. The non-interacting Hamiltonian \textcolor{black}{is given by}
\begin{equation}
        \mathcal{H}^{\chi,\sigma}_0(\mathbf{k})=
    \begin{pmatrix}
        \mathcal{H}^{1,\chi}(\mathbf{k}) & \mathcal{U}^{\chi,\dagger} & \mathbb{0} \\
        \mathcal{U}^{\chi} & \mathcal{H}^{2,\chi}(\mathbf{k}) & \mathcal{U}^{\chi} \\
        \mathbb{0} & \mathcal{U}^{\chi,\dagger} &  \mathcal{H}^{3,\chi}(\mathbf{k}) 
    \end{pmatrix},
\label{eq:H_cont_tTLG}
\end{equation}
\textcolor{black}{where} $\mathcal{H}^{3,\chi}(\mathbf{k})$ has the same form as $\mathcal{H}^{1,\chi}(\mathbf{k})$, \textcolor{black}{owing to the mirror symmetry}, and \textcolor{black}{$\mathcal{H}^{2,\chi}(\mathbf{k})$ remains unchanged from tBLG.}

\textcolor{black}{The Hartree contribution to the Hamiltonian for tTLG is analogous to Eq.~\eqref{eq:H_hart}, since $\exp(-2d|\textbf{G}|) \approx 1$. However, as the inner layer is distinct from the outer layers, a crystal field can lower the energy of the middle layer~\cite{Fischer_unconv_supercond_2022}, which motivates including $\delta \rho_0^l$. As we only investigate tTLG at charge neutrality here, for simplicity, we neglect the Hartree contribution to the Hamiltonian. }

\textcolor{black}{Analogously, the} Hubbard Hamiltonian is given by
\begin{equation}
\begin{aligned}
   \mathcal{H}^{U,l}_{N,M} = & \delta^l_{0} \mathcal{S} \delta^{(2)}(\mathbf{G}_{N}-\mathbf{G}_{M}) +\\ 
    & \delta^l_{1} \mathcal{S}\sum_{\mathbf{G}_\star}\delta^{(2)}(\mathbf{G}_{N}-\mathbf{G}_{M}-\mathbf{G}_{\star}).\\ 
\end{aligned}
\label{eq:H_Hub_tTLG}
\end{equation}
\textcolor{black}{From the symmetry of tLTG}, $\delta^3_{0/1}=\delta^1_{0/1}=\delta^o_{0/1}$ and $\delta^2_{0/1}=\delta^i_{0/1}$ for the other orders. Similar to tBLG, details of the exact form of the \textcolor{black}{magnetic orders} will be given in the Results section~\ref{sec:results}. 

Having defined these parameters, the full spin-valley-polarized Hamiltonian \textcolor{black}{of tLTG at charge neutrality}, incorporating \textcolor{black}{the non-interacting Hamiltonian} and short-ranged Hubbard interactions, is 
\begin{multline}
\mathcal{H}^{\chi,\sigma}(\mathbf{k}) = \mathcal{H}_0^{\chi,\sigma} + 
\begin{pmatrix}
\mathcal H^{U,\sigma,o} & \mathbb 0 & \mathbb 0 \\
\mathbb 0 & \mathcal H^{U,\sigma,i} & \mathbb 0 \\
\mathbb 0 & \mathbb 0 & \mathcal H^{U,\sigma,o} \\
\end{pmatrix}.
\label{eq:H_cont_fin_tTLG}
\end{multline}
Obtaining the eigenstates \textcolor{black}{of this Hamiltonian} allows us to compute the \textcolor{black}{order parameters $\delta^o_{0},\delta^o_{1},\delta^i_{0},\delta^i_{1}$, analogously to the twisted bilayer case. We use the same self-consistency criteria for these calculations.} 

\section{Results}\label{sec:results}

\subsection{Twisted Bilayer Graphene}

\subsubsection{Leading Instabilities}

\begin{figure}
    \centering
    \begin{minipage}[b]{0.49\textwidth}
    \sublabel{a}{FM}
    \includegraphics[width=\textwidth]{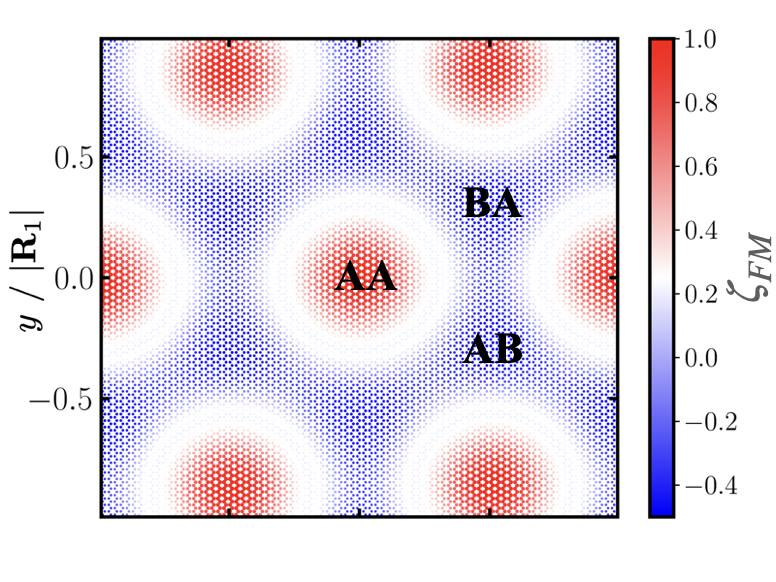}
\end{minipage}
\begin{minipage}[b]{0.49\textwidth}
    \sublabel{b}{MAFM}
    \includegraphics[width=\textwidth]{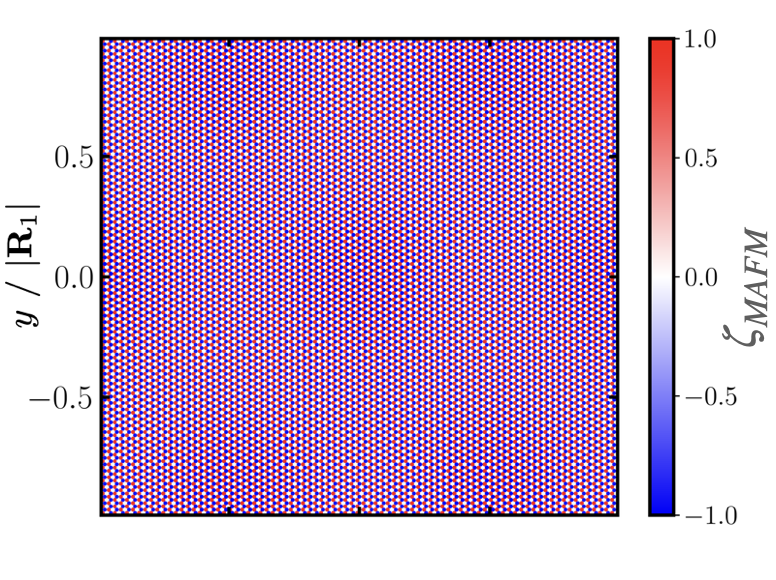}
\end{minipage}\\
\begin{minipage}[b]{0.49\textwidth}
    \sublabel{c}{NAFM}
    \includegraphics[width=\textwidth]{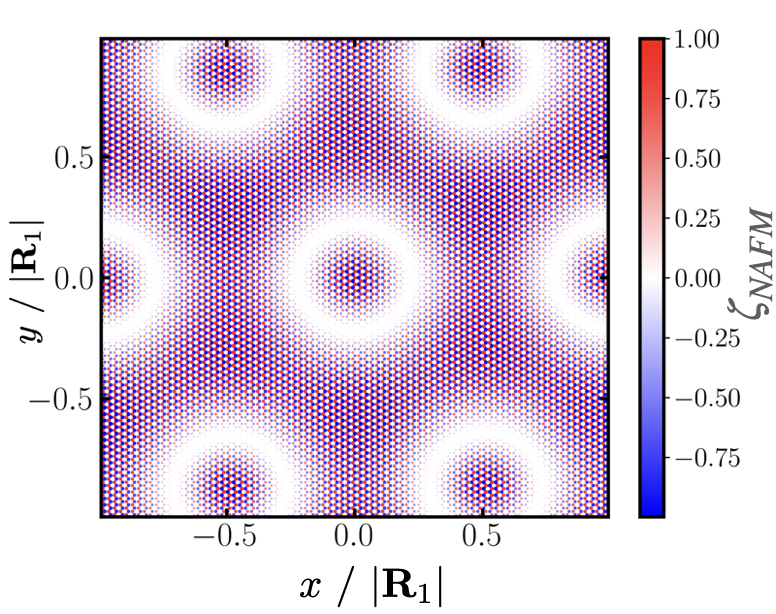}
\end{minipage}
    \caption{Magnetic instabilities obtained from atomistic RPA calculations. The ferromagnetic order (FM), moir\'e modulated antiferromagnetic (MAFM), and nodal antiferromagnetic (NAFM) orders are shown in a), b) and c), respectively. Note these graphs are obtained from the eigenvectors of these calculations\textcolor{black}{, corresponding to the most negative eignevalues}, where the largest positive value was chosen to be 1.}
    \label{fig:RPA_tBLG}
\end{figure}

The leading magnetic instabilities can be obtained \textcolor{black}{from atomistic RPA calculations at $\vet{q}=0$}~\cite{Klebl_mag_order_tBLG_2019,Goodwin_mag_tBLG_2021}. Details of \textcolor{black}{these calculations} can be found in the papers by Klebl~\textit{et al.}~\cite{Klebl_mag_order_tBLG_2019,Goodwin_mag_tBLG_2021}. The eigenvectors \textcolor{black}{of these RPA calculations characterize the type and form of} these magnetic instabilities, $\zeta_i$. The values of which \textcolor{black}{are proportional to} the spin-polarized electron density, $\zeta \propto  n_{\uparrow}-n_{\downarrow}$, where $n_{\uparrow\downarrow}$ \textcolor{black}{is the electron density of each spin}. The ferromagnetic order (FM), moir\'e modulated antiferromagnetic (MAFM), and nodal antiferromagnetic (NAFM) orders were found to be the most prevalent in magic-angle tBLG~\cite{Klebl_mag_order_tBLG_2019,Goodwin_mag_tBLG_2021}, which can be seen in Fig.~\ref{fig:RPA_tBLG}.

As seen in Fig.~\ref{fig:RPA_tBLG}, and discussed in Refs.~\cite{Klebl_mag_order_tBLG_2019,Goodwin_mag_tBLG_2021}, the FM order is characterized by a spin-polarized electron density peaked in the AA regions \textcolor{black}{of the moir\'e unit cell}. \textcolor{black}{This FM order} approximately follows a cosine series \textcolor{black}{with} the first reciprocal lattice vectors \textcolor{black}{and} a constant shift. As the AA region is centered at the origin, \textcolor{black}{we can} approximate the FM order by  
\begin{equation}
    \zeta_\mathrm{FM} = \zeta_{0} + \zeta_{1}\sum_{i}\cos(\mathbf{G}_{i}\mathbf{r}),
    \label{eq:zeta_FM}
\end{equation}
\noindent where $\mathbf{G}_{i}$ are the moir\'e reciprocal lattice vectors ($\mathbf{G}_{1}$, $\mathbf{G}_{2}$, and $\mathbf{G}_{1} + \mathbf{G}_{2}$), $\zeta_0$ is a parameter which characterizes the constant \textcolor{black}{contribution} spin-polarized electron density in the moir\'e unit cell, and $\zeta_1$ represents the moir\'e-scale modulation of the FM order. \textcolor{black}{Since} for the FM instability \textcolor{black}{the sublattice polarization is the same}, the sublattice-coupling matrix $\mathcal{S}$ is taken to be the identity matrix. 

For the MAFM order, there is a similar \textcolor{black}{moir\'e scale spin} structure, but the value of $\zeta$ changes sign between each sublattice, implying $\mathcal{S} \propto \sigma_z$. Therefore, the MAFM order can be approximately characterised by 
\begin{equation}
    \zeta_\mathrm{MAFM} = \zeta_{0} + \zeta_{1}\sum_{i}\cos(\mathbf{G}_{i}\mathbf{r}),
    \label{eq:zeta_MAFM}
\end{equation}
\noindent where $\zeta_0$ and $\zeta_1$ have opposite sign for each sublattice, but the same sign within each sublattice on different layers. 

Finally, the NAFM is analogous to the MAFM case, but without a constant-shift term, which means it can be approximated as
\begin{equation}
    \zeta_\mathrm{NAFM} = \zeta_{1}\sum_{i}\cos(\mathbf{G}_{i}\mathbf{r}).
    \label{eq:zeta_NAFM}
\end{equation}
\noindent Again $\zeta_1$ has the opposite sign for each sublattice, meaning $\mathcal{S} \propto \sigma_z$. 

\subsubsection{Regions of Stability}
\label{sec:tBLG_region_stab}

We included these leading magnetic instabilities, from the atomistic calculations in the previous section, in the continuum model and solved the resulting Hamiltonian self-consistently (see Sec.~\ref{sec:methods}). To perform these calculations, we need to choose a value of the on-site Hubbard parameter in the continuum model. \textcolor{black}{In the atomistic model, the value of bare $U_z$ for p$_z$ orbitals is known to be $\approx 10$~eV\cite{SECI,OHP}, which can be approximated by $U_z \approx e^2/4\pi\epsilon a_0$~\cite{EE}. Following Guinea and Walet~\cite{EE}, these p$_z$ values need to be projected onto the moir\'e wavefunctions, which gives the moir\'e Hubbard parameter to be $U \approx (U_z N^{-2})N \approx e^2a_0/4\pi\epsilon L^2$, where $N \approx (L/a_0)^2$ is the number of atoms in the moir\'e unit cell. Since the lattice parameter of graphene is $\sim 2$\AA and the moir\'e scale is $\sim 100$\AA, the value of $U \sim 1$~meV. In comparison, the Hartree interaction is $L/a_0$ larger, which dominates over the Hubbard interaction, and motivated neglecting the Hubbard interaction term previously~\cite{EE}. However, an energy scale of $\sim 1$~meV is still similar to the bandwidth of tBLG at the magic angle, which means these interactions can still be important~\cite{Alejandro23}. In this section, we} choose $U=2$~meV to solve our system of equations self-consistently, which is comparable to the values used by Jimeno-Pozo \textit{et al.}~\cite{Alejandro23}. 

\begin{figure*}[htbp!]
\centering
\begin{minipage}[b]{0.325\textwidth}
    \sublabel{a}{MAFM}
    \includegraphics[width=\textwidth]{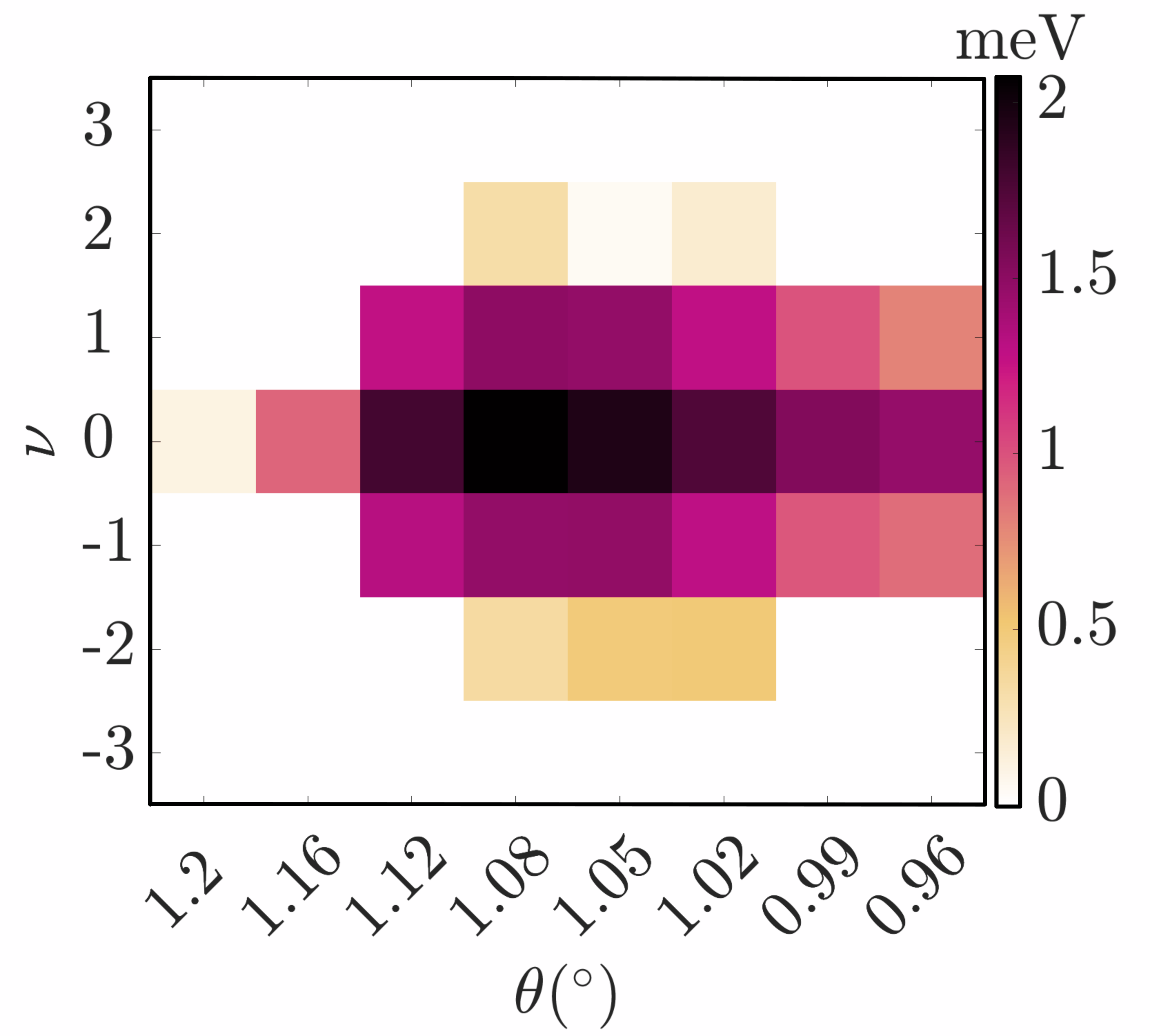}
\end{minipage}
\begin{minipage}[b]{0.325\textwidth}
    \sublabel{b}{NAFM}
    \includegraphics[width=\textwidth]{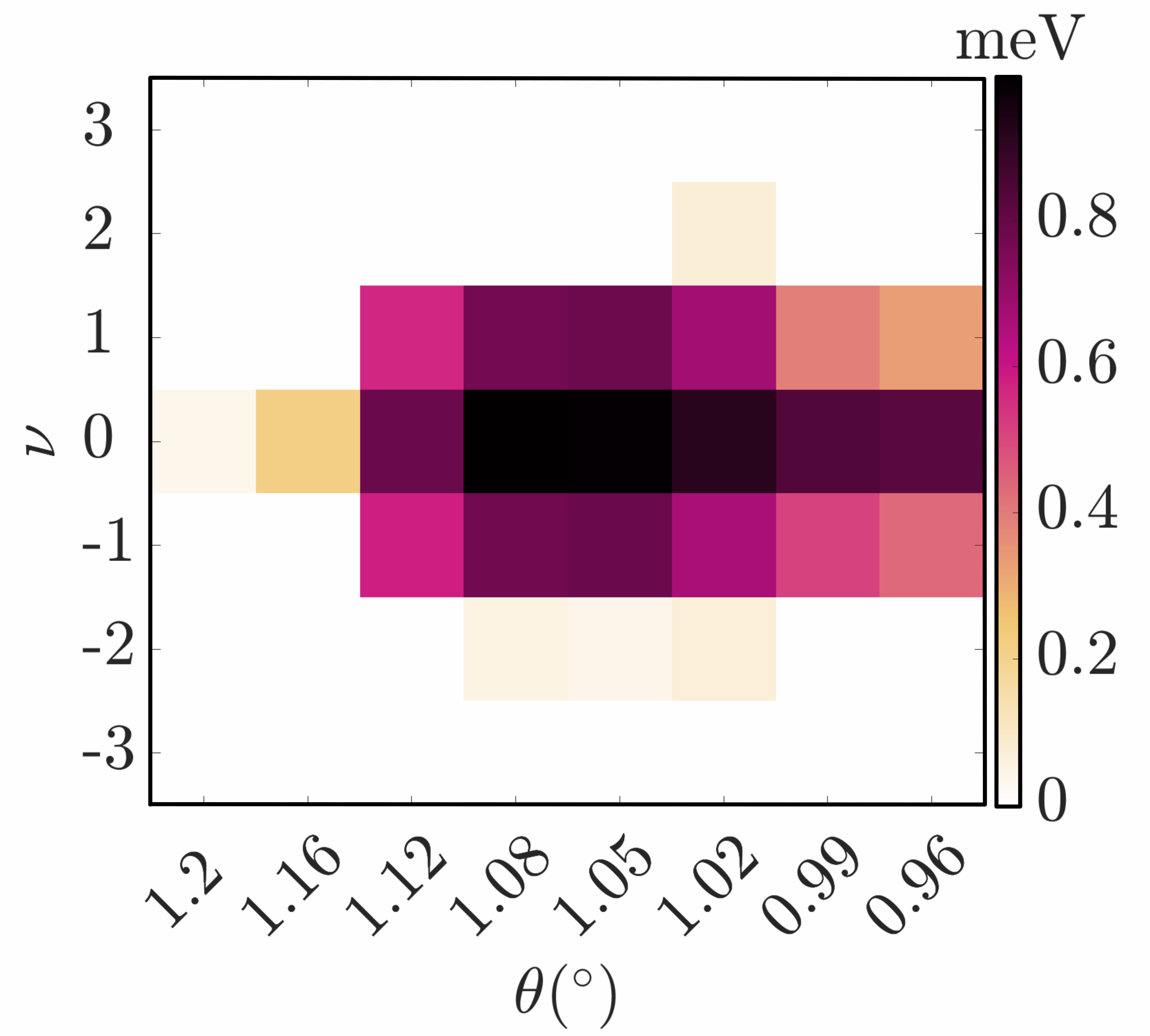}
\end{minipage}
\begin{minipage}[b]{0.325\textwidth}
    \sublabel{c}{FM}
    \includegraphics[width=\textwidth]{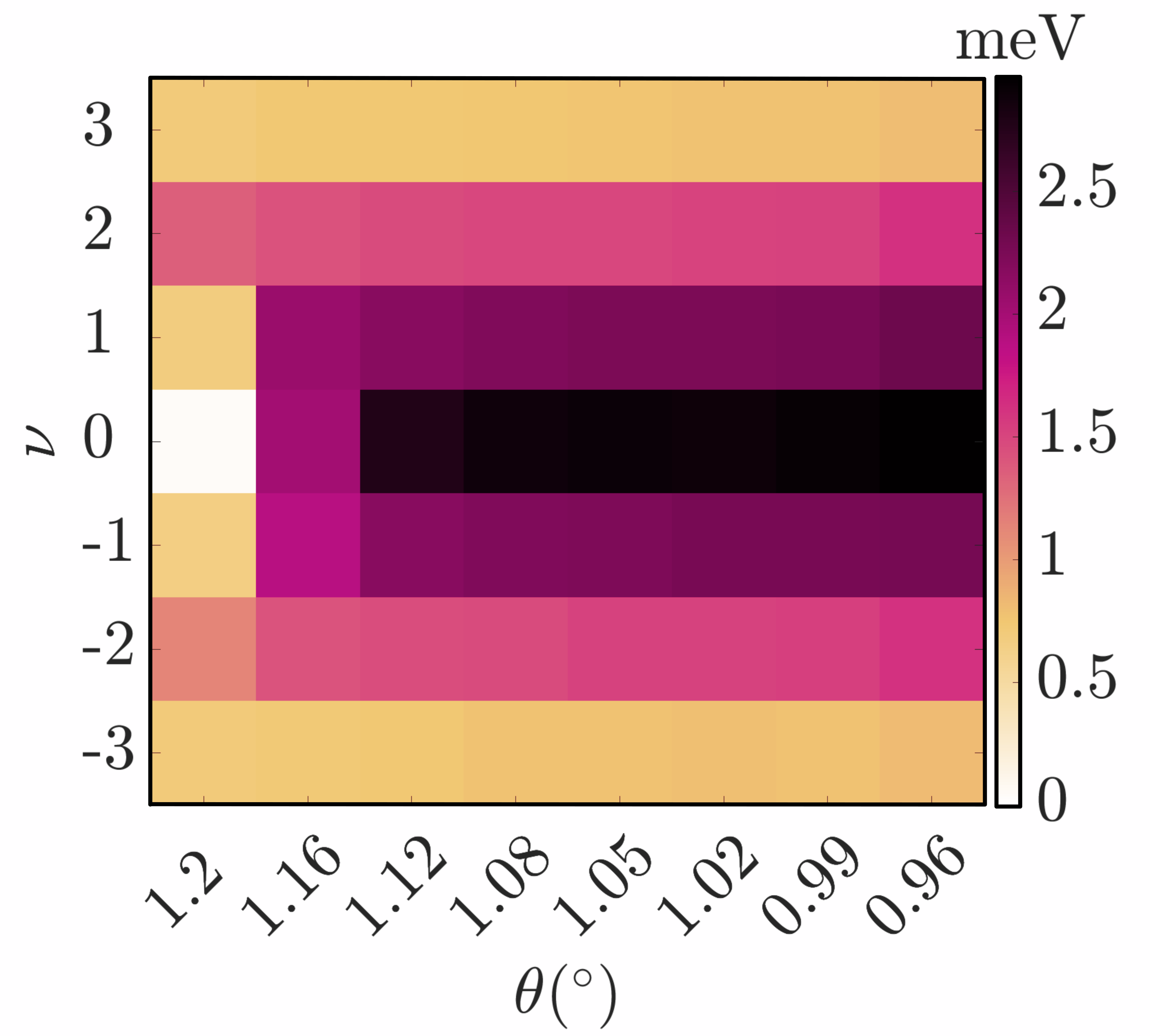}
\end{minipage}
\par\bigskip
\begin{minipage}[b]{0.325\textwidth}
    \sublabel{d}{}
    \includegraphics[width=\textwidth]{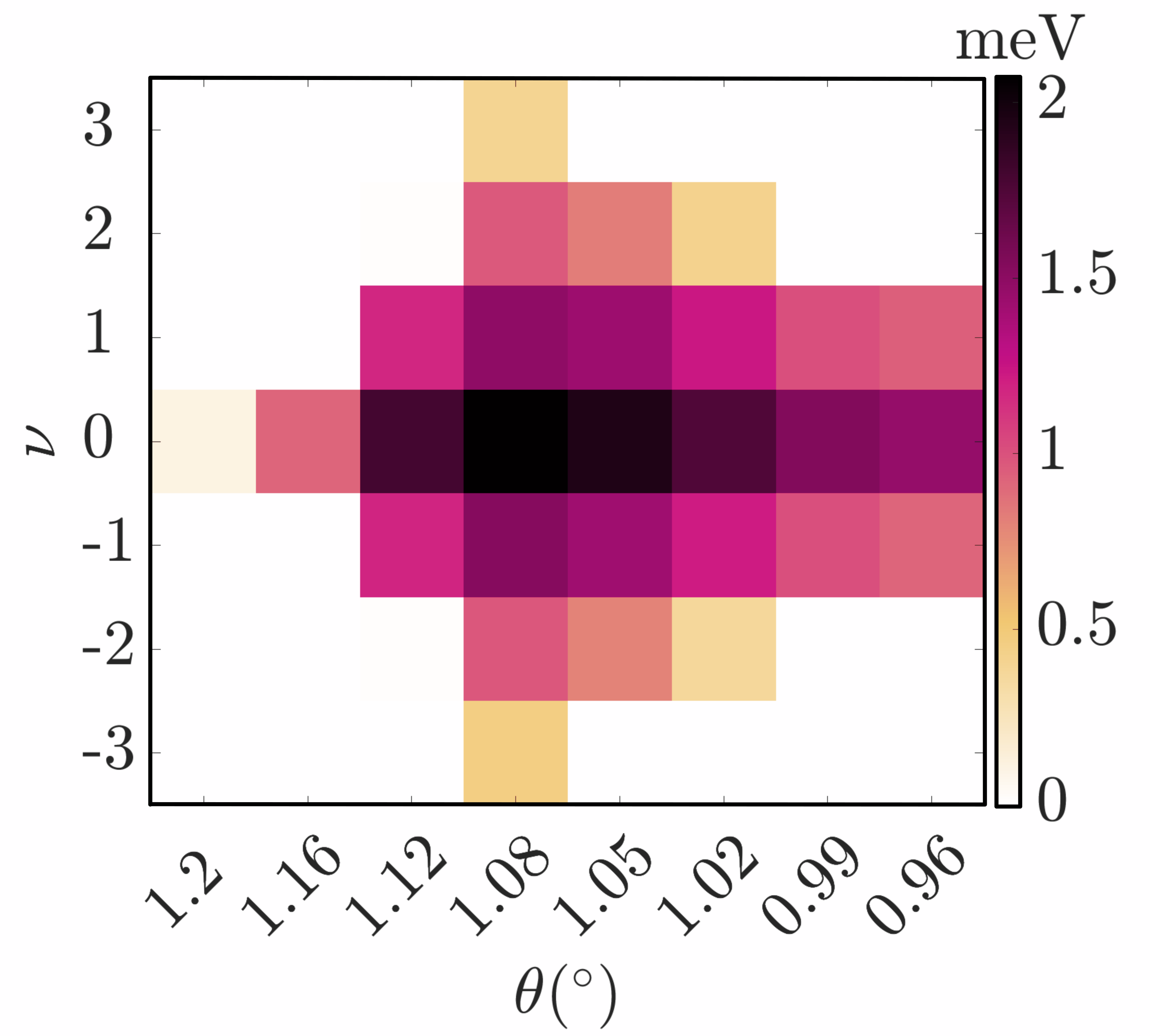}
\end{minipage}
\begin{minipage}[b]{0.325\textwidth}
    \sublabel{e}{}
    \includegraphics[width=\textwidth]{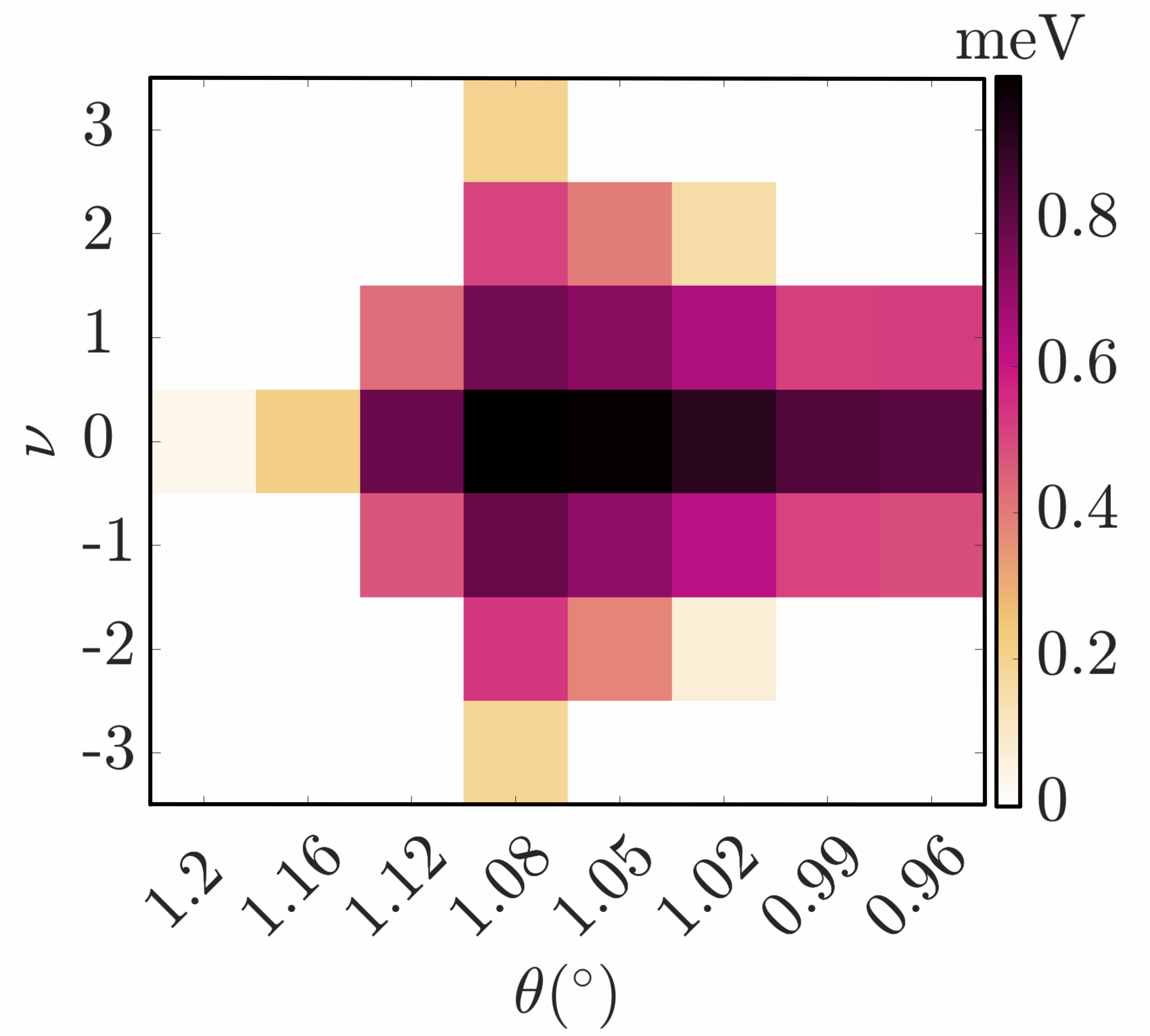}
\end{minipage}
\begin{minipage}[b]{0.325\textwidth}
    \sublabel{f}{}
    \includegraphics[width=\textwidth]{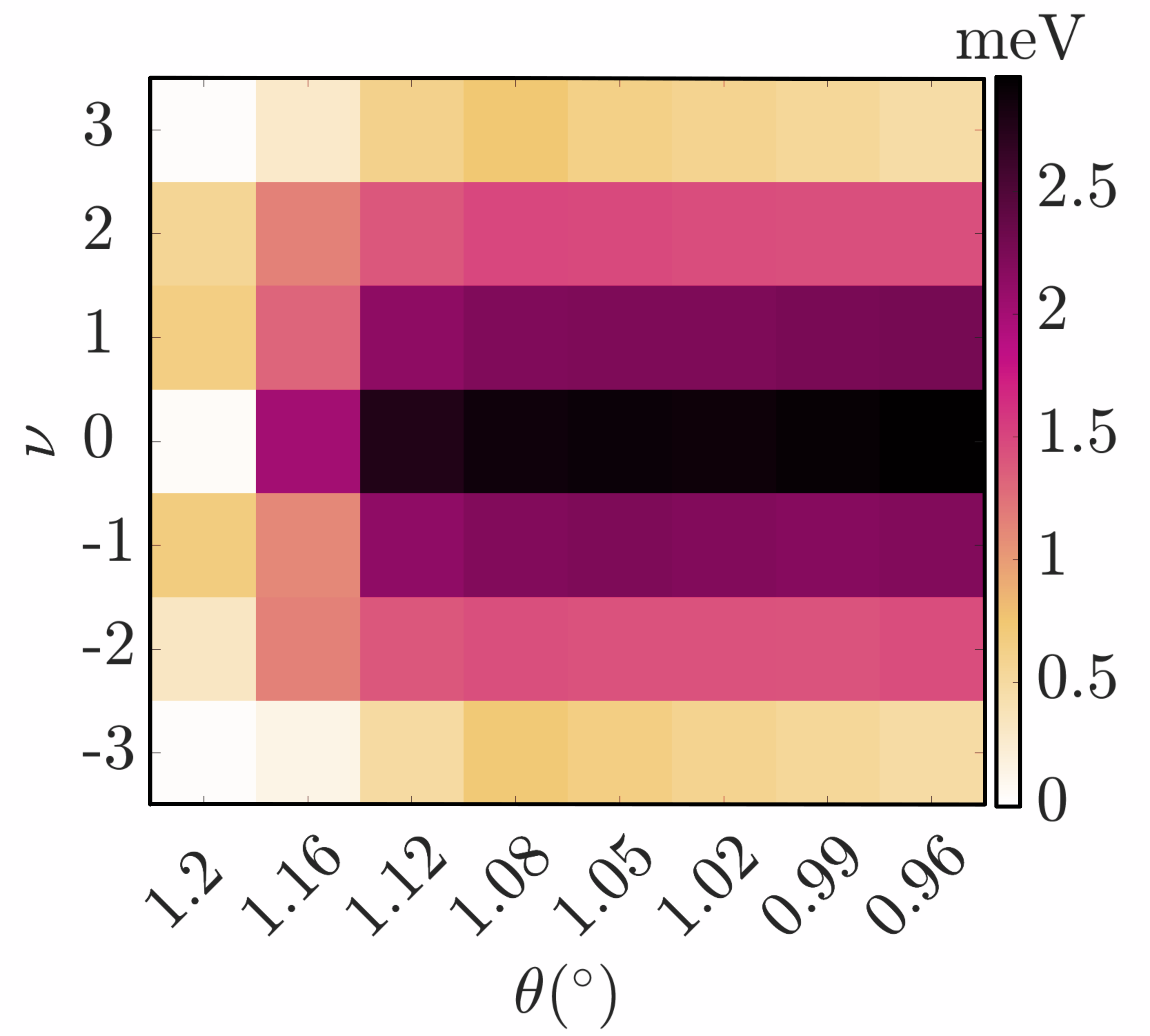}
\end{minipage}
\caption{Magnitudes of Hubbard potential order parameters $|\delta_0| + |\delta_1|$, for the \textcolor{black}{considered} magnetic ground states, \textcolor{black}{to establish where these phases can exist at twist angle ($\theta$) close to the magic angle and for integer doping levels within the flat bands ($\nu$)}. \textcolor{black}{With Hartree interactions, we show results for MAFM in} (a), NAFM in (b), and FM in (c). Results without Hartree interactions are shown in (d) for MAFM, (e) for NAFM, and (d) for FM. $U =2$~meV throughout. We note that $\delta_0=0$ for the NAFM order.}
\label{fig:Vtots_tBLG}
\end{figure*}

To quantify the stability of the magnetic orders, the magnitudes of order parameters $|\delta_0|+|\delta_1|$ are displayed in Fig.~\ref{fig:Vtots_tBLG} as a function of doping level ($\nu$) and twist angle ($\theta$), with and without long-range Coulomb interactions. The cutoff for \textcolor{black}{considering a magnetic order as stable} is set at $|\delta_0|+|\delta_1|\leq5\times10^{-2}$~meV, \textcolor{black}{ which is comparable to our convergence criteria,} below which the Hubbard potentials are too weak to induce any observable modifications to the band structures.

In Fig.~\ref{fig:Vtots_tBLG}(a) \textcolor{black}{we show results for the total order parameter ($|\delta_0| + |\delta_1|$) of MAFM, which indicates if this magnetic order survives at the considered twist angle and doping level,} with account of Hartree interactions. The MAFM magnetic order \textcolor{black}{strongly} persists from $\theta = 1.12 \degree$ to $\theta = 0.96 \degree$ \textcolor{black}{inside the doping range of} $\nu = \pm1$. Notably, the symmetry broken state is most strongly stabilized near charge neutrality at \textcolor{black}{the magic-angle} $\theta=1.08 \degree$, with a total order parameter of $2.1$~meV, away from which the order parameters are suppressed. Upon doping in the range of $\theta=1.08\degree-1.02~\degree$, stable MAFM order persists up to $\nu=\pm2$, albeit being reduced by an order of magnitude to $\sim$0.1~meV at $\nu=\pm2$ for $\theta=1.08\degree-1.02~\degree$. For twist angles outside of this window the order parameter decays more rapidly as the system is doped, i.e., more doping levels near the magic-angle host magnetic order, while the twist angles further from the magic angle only host magnetic order near charge neutrality.

The order parameter for NAFM \textcolor{black}{with Hartree interactions}, as shown in Fig.~\ref{fig:Vtots_tBLG}(b), exhibits a similar \textcolor{black}{trends} as function of the twist angle and doping as the MAFM case. The magnitude \textcolor{black}{of the order parameter for NAFM relative to MAFM is} reduced by approximately $1/2$ because NAFM order does not have $\delta_0$. Similar to MAFM, NAFM magnetic order is most stable near charge neutrality at the magic angle\textcolor{black}{, with it strongly decaying in the doping variable, hardly existing outside of $\nu = \pm1$, but weakly decaying with twist angle away from the magic angle.}

The FM order \textcolor{black}{with Hartree interactions}, shown in Fig.~\ref{fig:Vtots_tBLG}(c), is the most stable at charge neutrality, with a total order parameter around 3~meV near the magic angle. The FM order is also suppressed upon doping the system, while being \textcolor{black}{relatively} constant over twist angles near the magic angle. Upon twisting to larger angles, the magnetic order is suppressed at charge neutrality first.  At $\theta=1.2 \degree$, the FM order is suppressed at charge neutrality, but stable FM orders are notably developed near $\nu=\pm2$, with a total order parameter amplitude of around 1.5~meV at $\nu=\pm2$. The $\theta=1.2 \degree$ twist angle is the only one we observed to have a non-monotonic doping dependence of the FM, whereas for all other angles the FM order parameter decays monotonically from charge neutrality. \textcolor{black}{At smaller angles than those considered here, it is expected the FM is again destabilized first at charge neutrality.}

The effect of long-ranged interactions is also investigated by considering the value of the magnetic order parameter when the Hartree interactions are not taken into account. In other words, we consider the stability of the magnetic order for different twist angles and doping levels when only short-ranged Hubbard interactions are present. The phase diagrams for these calculations are shown in Figs.~\ref{fig:Vtots_tBLG}(d-f). 

The antiferromagnetic orders, MAFM [Fig.~\ref{fig:Vtots_tBLG}(d)] and NAFM [Fig.~\ref{fig:Vtots_tBLG}(e)], have clear differences in the absence of Hartree interactions. The phase diagrams are \textcolor{black}{slightly} more electron-hole symmetric compared to when Hartree interactions are taken into account. Furthermore, the magnetic order parameter exists over a wider range of doping levels, and is significant at $\nu=\pm3$ \textcolor{black}{for the magic angle} $\theta=1.08 \degree$. Otherwise, the qualitative variations remain similar at other twist angles and dopings \textcolor{black}{compared to when long-range Coulomb interactions are taken into account. We don't notice any large differences in the magnitudes of the order parameters between these cases}.

For the FM order \textcolor{black}{without Hartree interactions}, as shown in Fig.~\ref{fig:Vtots_tBLG}(f), the qualitative variation and magnitude of the order parameters are similar to that in the presence of Hartree interaction. However, the reduction of the order parameters is more rapid with doping levels away from charge neutrality, in contrast to the anti-ferromagnetic order. For example, at $\theta=1.16 \degree$, although the stable orders are still centred around charge neutrality, the magnitude at $\nu=\pm3$ is significantly reduced. \textcolor{black}{Moreover, at $\theta=1.2 \degree$, FM order is only stable at $\nu = \pm1$ and $\pm2$.}

\textcolor{black}{Overall}, comparing Fig.~\ref{fig:Vtots_tBLG}(a) and \ref{fig:Vtots_tBLG}(d), and \ref{fig:Vtots_tBLG}(b) and \ref{fig:Vtots_tBLG}(e), we find that the magnetic orders M(N)AFM are enhanced by the long-ranged Hartree interactions at $\nu=\pm 1$ near the magic angle. Beyond this doping level, Hartree interactions suppress magnetic order. On the contrary, comparing Figure \ref{fig:Vtots_tBLG}(c) and \ref{fig:Vtots_tBLG}(f), the FM order is always enhanced by Hartree interactions. Such differences lie in the interplay between short and long-ranged interactions, which also underpins the distortions of the band structures. 

\subsubsection{Band Structure}

\begin{figure*}[htbp!]
\centering
\begin{minipage}[b]{0.45\textwidth}
    \sublabel{a}{NONE}
    \includegraphics[width=\textwidth]{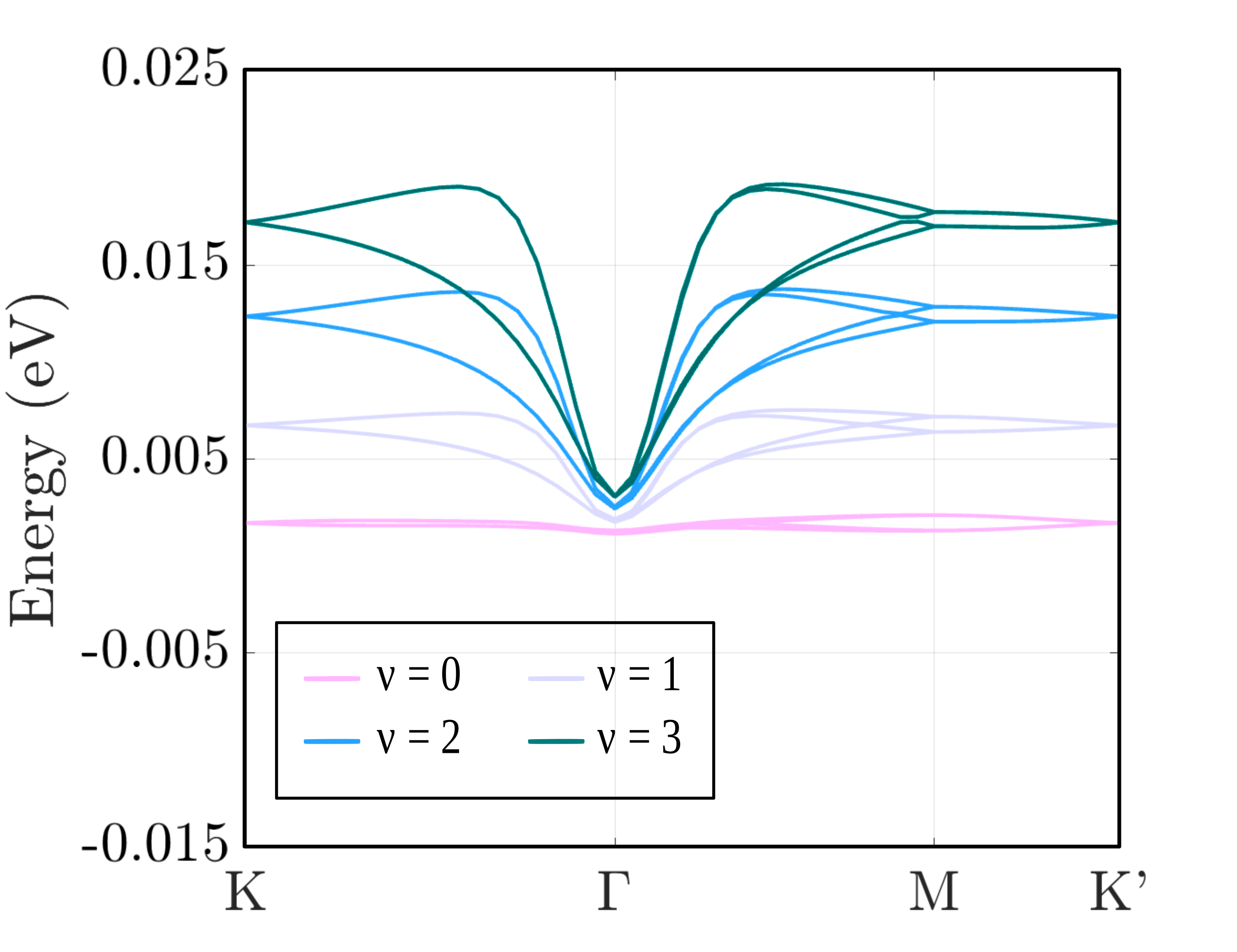}
\end{minipage}
\begin{minipage}[b]{0.45\textwidth}
    \sublabel{b}{MAFM}
    \includegraphics[width=\textwidth]{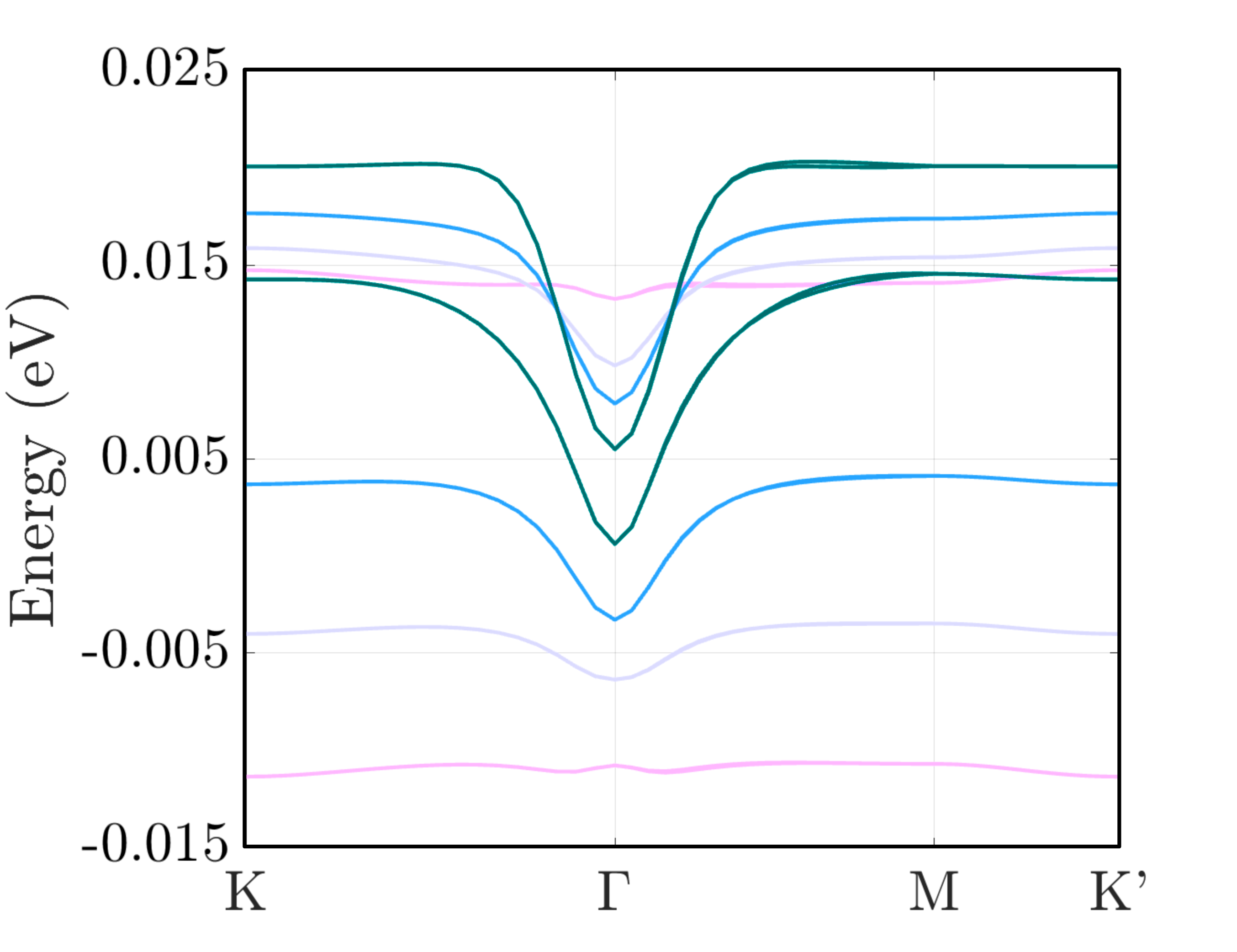}
\end{minipage}\\
\begin{minipage}[b]{0.45\textwidth}
    \sublabel{c}{NAFM}
    \includegraphics[width=\textwidth]{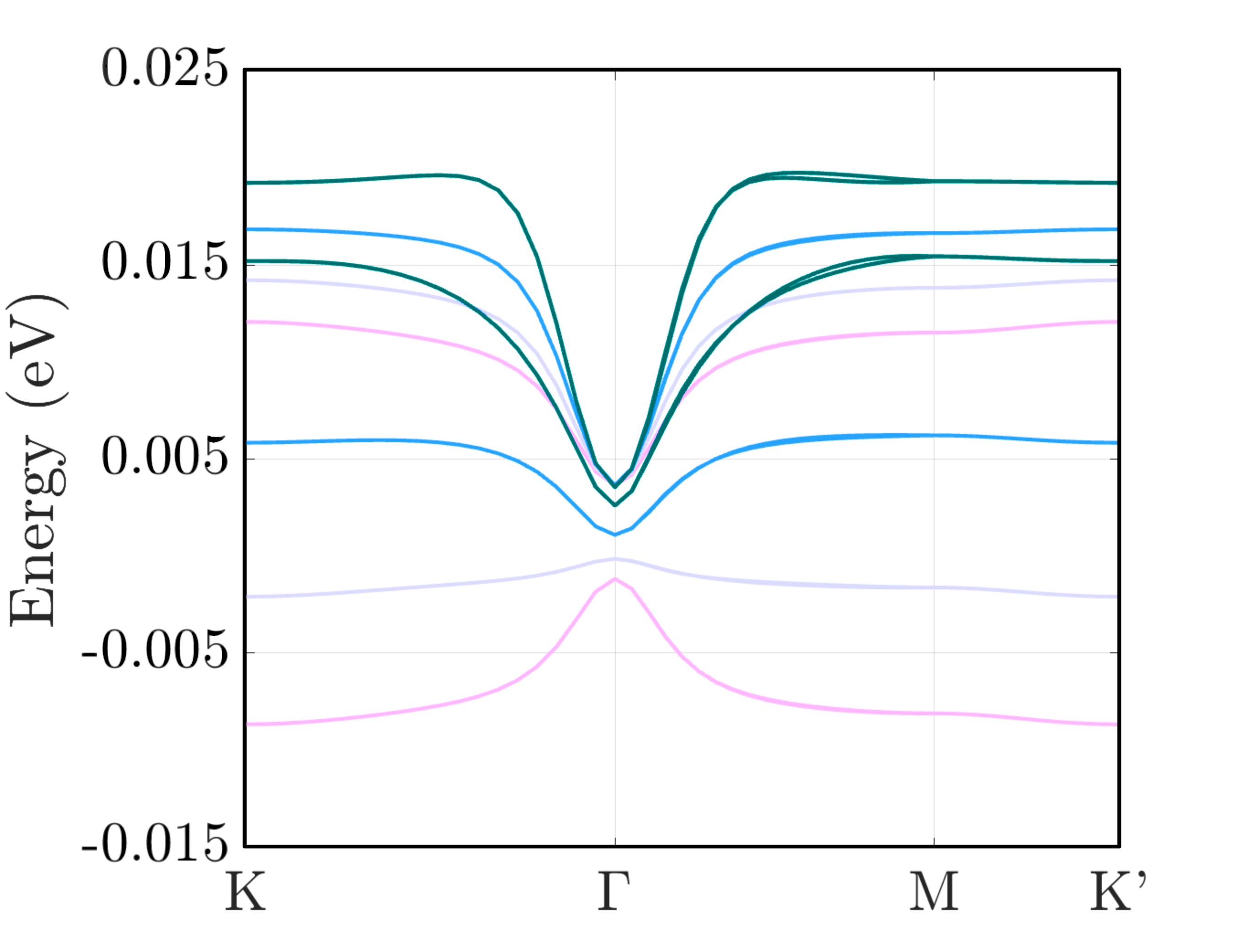}
\end{minipage}
\begin{minipage}[b]{0.45\textwidth}
    \sublabel{d}{FM}
    \includegraphics[width=\textwidth]{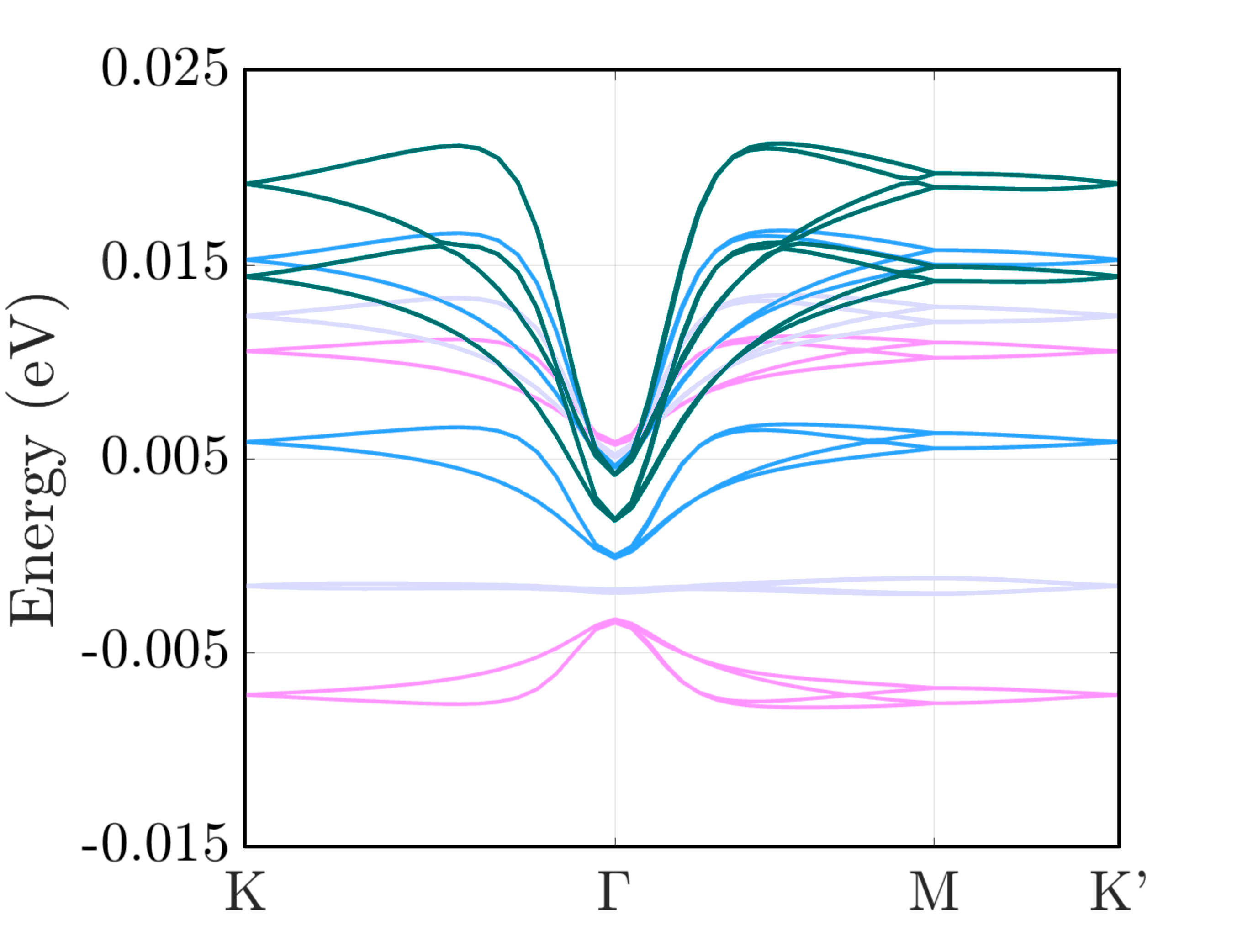}
\end{minipage}
\caption{The self-consistent band structures at 1.08\degree for different fillings $\nu$ for (a): Hartree corrections only, (b): MAFM, (c): NAFM, and (d): FM. We note that a large Hubbard parameter ($U=4$~meV) was chosen to demonstrate the effects of gap opening on Hartree-corrected bands. Color represents different filling factors: light pink for $\nu=0$; light blue for $\nu=1$; blue for $\nu=2$ and dark green for $\nu=3$.}
\label{fig:BS_tBLG}
\end{figure*}

The advantage of the approach developed here, unlike the atomistic RPA approach~\cite{Klebl_mag_order_tBLG_2019} and the perturbative continuum approach~\cite{Alejandro23}, is that self-consistent quasi-particle band structures can also be investigated. In this section, we report these for magic-angle tBLG for the leading instabilities, as a function of doping with the inclusion of long-range Hartree interactions \textcolor{black}{, as seen in Fig.~\ref{fig:BS_tBLG}}. We restrict our analysis here to only electron-doped systems, as hole-doping is qualitatively similar (albeit with band distortions in the opposite direction of energy)~\cite{PHD_4}. \textcolor{black}{Moreover, we use a value of $U = 4$~meV to stabilize the magnetic orders at all doping levels to make it more clear how gaps at high symmetry points vary with doping level, as summarized in Tab.~\ref{tab:band_gaps}. In the Supporting Information (SI) we show additional results for the band structures with $U = 2$~meV, with their band gaps also being reported in Tab.~\ref{tab:band_gaps}. }

\begin{table}
    \centering
\begin{tabular}{llrrrr}
    \hline \hline
    $U$ & \textbf{Order} & $\nu = 0$ & $\nu = 1$ & $\nu = 2$ & $\nu = 3$  \\
    \hline 
    4&\textbf{FM}  & 18 & 14 & 9  & 5 \\
    &\textbf{MAFM} & 26 & 20 & 14 & 6 \\
    &\textbf{NAFM} & 21 & 16 & 11 & 4 \\
    \hline
    2&\textbf{FM}  & 9 & 7 & 5 & 2 \\
    &\textbf{MAFM} & 4 & 4 & 1 & 0 \\
    &\textbf{NAFM} & 4 & 3 & 0 & 0 \\
    \hline \hline
    \end{tabular}
    \caption{\textcolor{black}{Values of the band gaps, in units of meV, at K/K' points for the AFM magnetic order (MAFM and NAFM), and values between the spin-split K/K' points of the FM order. These are shown for different values of $U$, in units of meV, and for the indicated doping levels at the studied twist angle of 1.08\degree.}}
    \label{tab:band_gaps}
\end{table}

The band structures without magnetic order but with Hartree interactions (note that when \textcolor{black}{Hartree interactions} are not taken into account, it is simply that the $\nu=0$ band structure \textcolor{black}{applies} for all doping levels, as the Hartree interactions in the model at charge neutrality vanish) are shown in Fig.~\ref{fig:BS_tBLG}(a) for reference. As electrons are added into the normal state of tBLG, this causes the K-points to increase in energy relative to the $\Gamma$-points, causing a sensitive dependence of the band structure on doping, and a pinning of the van Hove singularities. At the magic angle, this causes the bands to become more dispersive, but at other twist angles these distortions cause doping-induced band flattening, in addition to the twist-angle induced band flattening. A detailed analysis of the band distortions induced by Hartree interaction is presented in references \cite{Cea_Hart_Pin_2019,Cea_Hart_Dist_2020,Cea2021,Goodwin2020}. 

The band structures for the MAFM and NAFM magnetic orders are shown in Fig.~\ref{fig:BS_tBLG}(b) and \ref{fig:BS_tBLG}(c), respectively. For both of these magnetic orders, a large gap is created at the K/K' points, owing to the sublattice symmetry breaking. For the charge neutral cases, the electronic bands are extremely flat over most of the Brillouin zone, \textcolor{black}{and the system is a narrow-gap correlated insulator.} The effect of Hartree interactions  \textcolor{black}{with doping} causes the bands to distort to higher energies, as was the case without magnetic order. For the other doping levels, the system becomes a metal, owing to partially filled dispersive bands. In each case, however, there remains a significant gap at the K/K' point, \textcolor{black}{albeit decreasing with increasing doping level away from charge neutrality,} with $\nu=3$ only having a small gap is present. \textcolor{black}{In Tab.~\ref{tab:band_gaps} we report the values of the gaps at the K/K' points for these magnetic orders.}

The band structures for FM order are shown in Fig.~\ref{fig:BS_tBLG}(d). At each doping level there exist two sets of bands around the Fermi energy, which can be attributed to the spin-polarisation of the electronic states, but the valley symmetry is not broken. At charge neutrality, one set of these bands is pushed lower in energy and the other set of bands are pushed to higher energies. Overall, at charge neutrality, a gap exists at the Fermi energy, which means the system is a FM insulator. \textcolor{black}{In addition to the spin-splitting of the bands, within each spin-polarised band there are significant distortions to the electronic structure analogous to Hartree interactions. For the spin-state that is lowered in energy, the states at the edge of the Brillouin zone are lower in energy than the states near the $\Gamma$ point. While for the band of the spin state that is higher in energy, the states at the edge of the Brillouin zone being further raised with respect to the $\Gamma$ point. Therefore, opposite distortions occur for each spin state. Moreover, while these distortions are analogous to Hartree interactions with doping, the opposite trends occur. Specifically, for the completely filled spin band the Hubbard interactions cause the states at the edge of the Brillouin zone to lower, while Hartree interactions for a completely filled band push these states higher in energy.}

For the other doping levels, the Hartree interactions \textcolor{black}{compete with} the band distortions \textcolor{black}{arising from the Hubbard interactions}. This causes the system to become an FM metal instead of being an insulator, as there is overlap of the spin-polarised bands or/and partially filled bands. \textcolor{black}{With increasing doping level, we know that the FM state monotonically weakens, see Fig.~\ref{fig:BS_tBLG}. This can clearly be seen from the shrinking energy separation of the spin-split Dirac cones at K/K'. In Tab.~\ref{tab:band_gaps} we also report the gaps between the spin-split Dirac cones at the K/K' points.}

\subsection{Twisted Trilayer Graphene}

\subsubsection{Leading Instabilities}

\begin{figure*}
    \centering
        \begin{minipage}[b]{0.45\textwidth}
    \sublabel{a}{FM (outside layer)}
    \includegraphics[width=\textwidth]{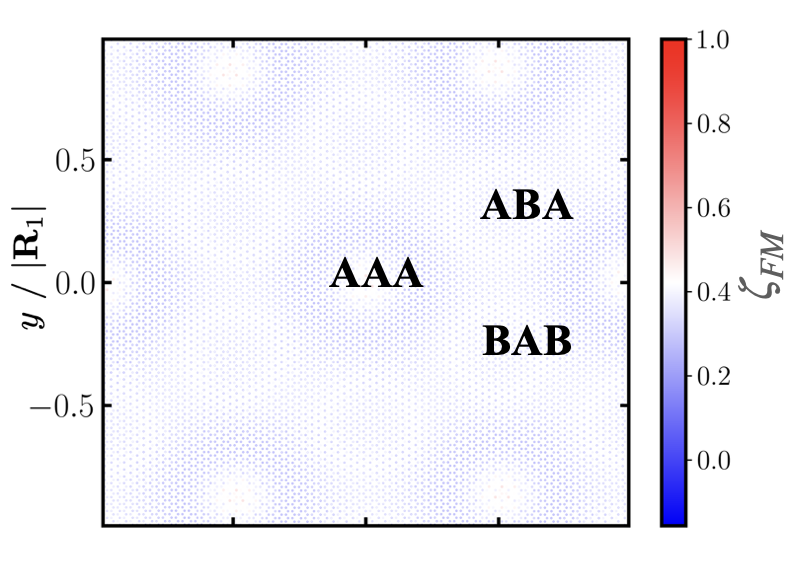}
    \end{minipage}
    \begin{minipage}[b]{0.40\textwidth}
        \sublabel{b}{MMAFM (outside layer)}
        \includegraphics[width=\textwidth]{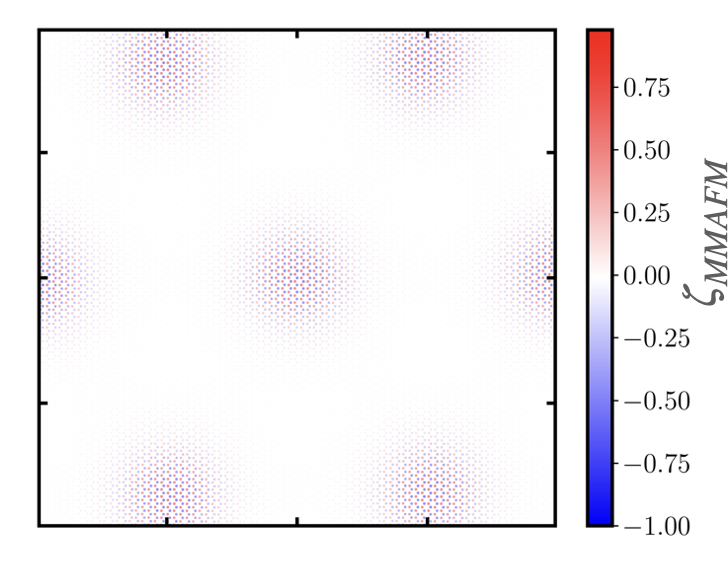}
    \end{minipage}
    \begin{minipage}[b]{0.45\textwidth}
        \sublabel{c}{FM (inside layer)}
        \includegraphics[width=\textwidth]{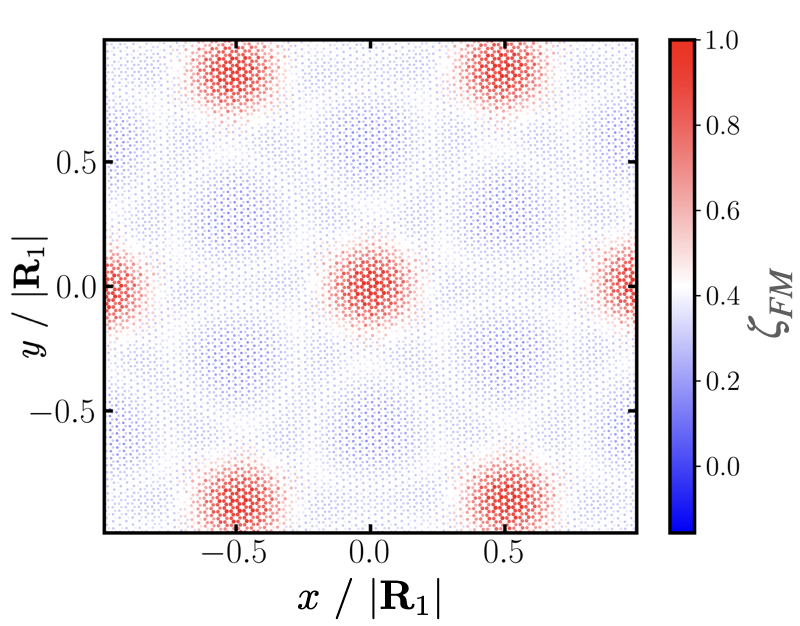}
    \end{minipage}
    \begin{minipage}[b]{0.39\textwidth}
        \sublabel{d}{MMAFM (inside layer)}
        \includegraphics[width=\textwidth]{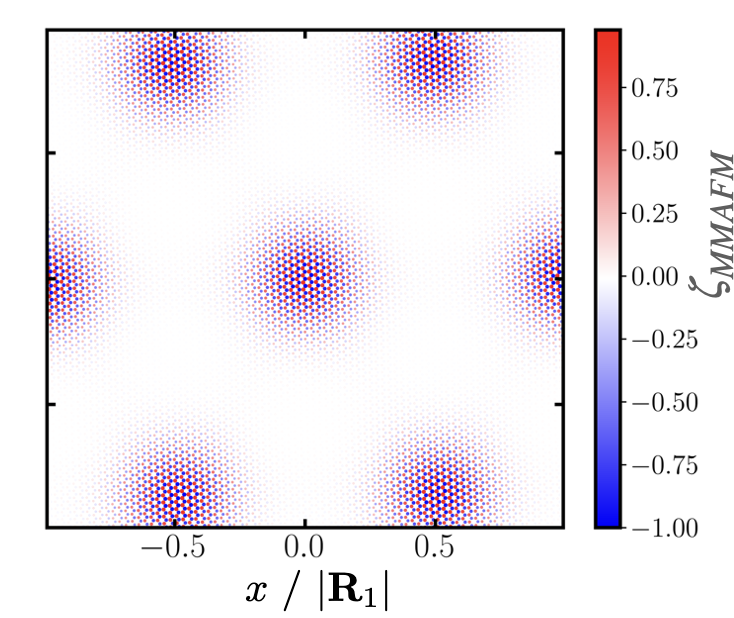}
    \end{minipage}
    \caption{Magnetic instabilities obtained from atomistic RPA calculations. \textcolor{black}{Here we only consider the} ferromagnetic order (FM) and MAFM/MAFM (MMAFM) order. \textcolor{black}{Since the outside and insider layers are inequivalent, we show them separately, and since the outside layers are equivalent we only show one of them. The outside and inside layers of FM are shown in a) and c), respectively, while the outside and inside layers of MMAFM are respectively in b) and d). }. Note these plots are from the eigenvectors of the RPA calculations, where the largest positive value was chosen to be $\pm$1. These values are proportional to the spin-polarised electron density.}
    \label{fig:RPA_ttlg}
\end{figure*}

We now turn our attention to the \textcolor{black}{mirror-symmetric} twisted trilayer graphene (tTLG). This system also exhibits a magic angle, but at a twist angle of $|\theta|=1.54\degree$~\cite{Fischer_unconv_supercond_2022}. For tTLG, we follow the same procedure for our calculations as tBLG~\cite{Fischer_unconv_supercond_2022}. First we found the leading instabilities from the RPA calculations, inspected these magnetic instabilities (\textcolor{black}{only some of which are} shown in Fig.~\ref{fig:RPA_ttlg}), constructed appropriate approximate forms for the order parameters and performed self-consistent continuum Hartree+U calculations. 

Similarly to tBLG, we find a ferromagnetic order (FM) in tTLG. As the inside and outside layer(s) of tTLG are inequivalent, we show an outside layer in  Fig.~\ref{fig:RPA_ttlg}a) and the inside layer in Fig.~\ref{fig:RPA_ttlg}c). We find the FM order is peaked mainly on the AAA regions of the inner layer~\cite{Fischer_unconv_supercond_2022}. Analogously to the bilayer, we can approximate this magnetic order with the form
\begin{equation}
    \zeta_{FM}^{(i)} = \zeta_{0}^{(i)} + \zeta_{1}^{(i)}\sum_{j}\cos(\mathbf{G}_{j}\mathbf{r}),
    \label{eq:zeta_FM_tLTG}
\end{equation}

\noindent where $(i)$ is now a layer index, as the order parameters can now vary between the layers. From the symmetry of the structure, the outside layers are equivalent, so we only need to consider the parameters to be distinct for outside vs. inside. Therefore, there are only four parameters for this magnetic order. The direction of such polarization is the same on the different sublattices, meaning that the sublattice-coupling $\mathcal{S}$ is taken to be the identity matrix for each layer.

Another candidate order is the MAFM/MAFM (MMAFM) magnetic order, where both the outer and inner layers are in a moir\'e-scale antiferromagnetic state (MAFM), as seen in Fig.~\ref{fig:RPA_ttlg}b) and d) for the outer and inner layers, respectively. This can be characterized by the following form
\begin{equation}
    \zeta_\mathrm{MMAFM}^{(i)} = \zeta_{0}^{(i)} + \zeta_{1}^{(i)}\sum_{j}\cos(\mathbf{G}_{j}\mathbf{r}).
    \label{eq:zeta_MMAFMtLTG}
\end{equation}

\noindent In this case, the spin polarization in each layer is identical, and similarly to the MAFM case for tBLG, there is an opposite sign for each sublattice, meaning that $\mathcal{S}$ is taken as the Pauli matrix $\sigma_z$. 

With the magnetic orders from the atomistic RPA transcribed into the continuum description, we solved for the magnetic orders self-consistently and obtained the band structures self-consistently at the magic angle of $1.54\degree$ in tTLG. Further details of how these magnetic orders are included in the Hamiltonian are given in the Methods section~\ref{sec:methods}. \textcolor{black}{Note we do not construct a full phase diagram for tTLG, as we did for tBLG, as a full RPA phase diagram for tTLG has not been computed. Moreover, the magnetic orders of tTLG can be a lot more varied than tBLG, but we only choose to study two here as a representative case to demonstrate our method can easily be extended to other moir\'e graphene multilayers. }

\subsubsection{Band structures}

\begin{figure*}[htbp!]
\centering
\begin{minipage}[b]{0.45\textwidth}
    \sublabel{a}{FM}
    \includegraphics[width=\textwidth]{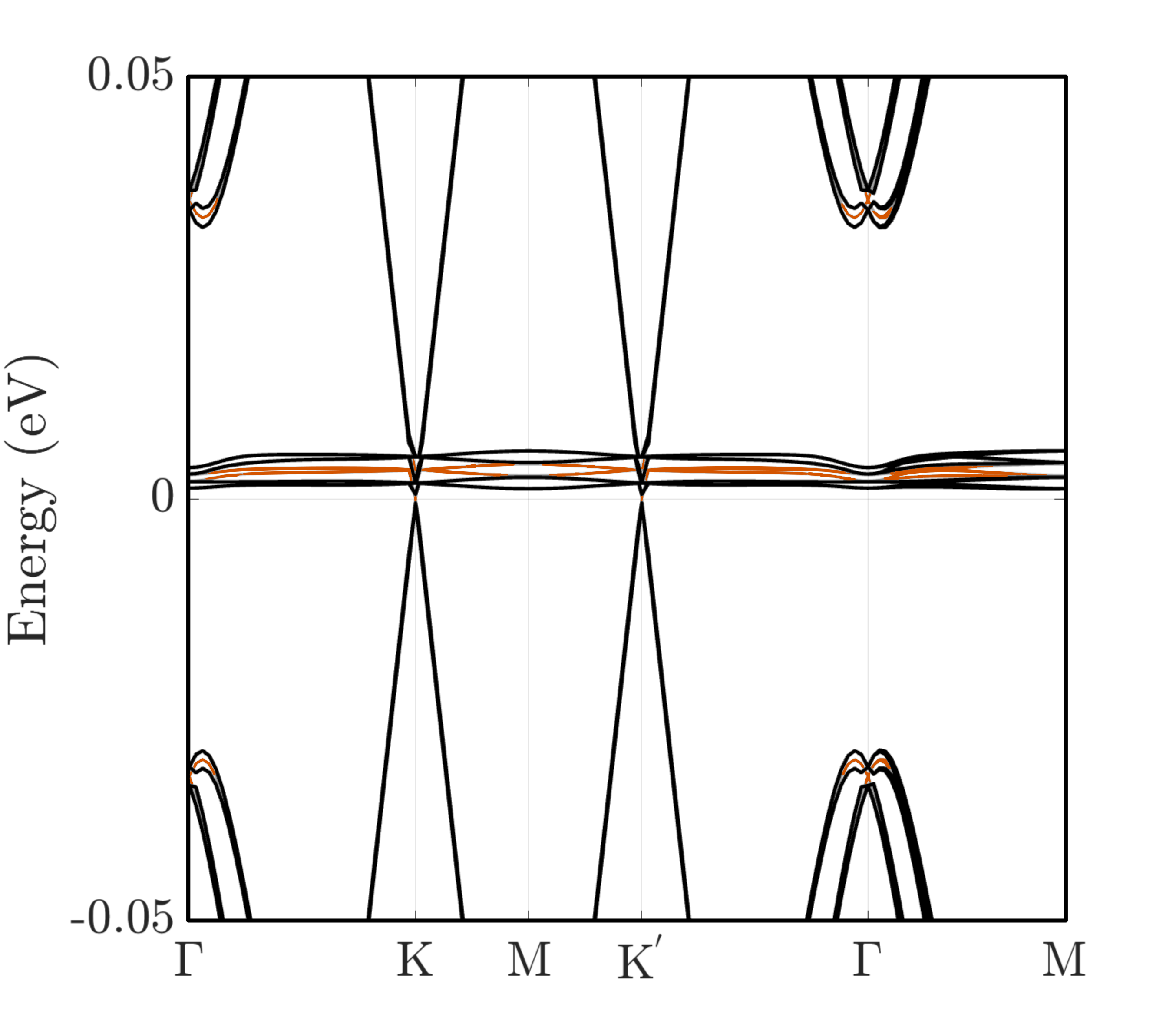}
\end{minipage}
\begin{minipage}[b]{0.45\textwidth}
    \sublabel{b}{MMAFM}
    \includegraphics[width=\textwidth]{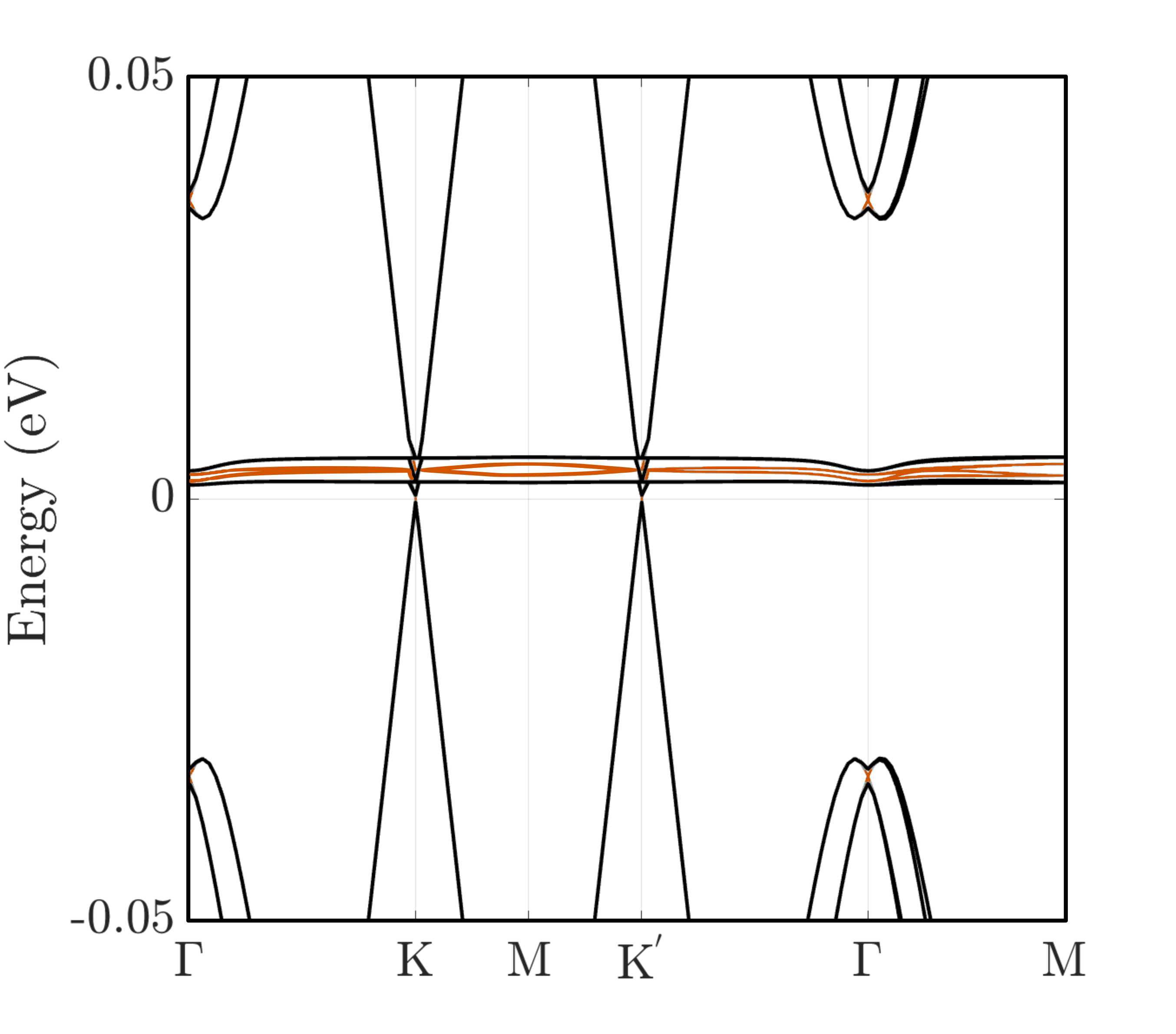}
\end{minipage}

\caption{Self-consistent band structures obtained of tTLG at charge neutrality at $\theta=1.54^{\circ}$ for (a) FM and (b) MMAFM. The band structure at charge neutrality without any Hubbard interactions is included in orange.}
\label{fig:BS_tTLG}
\end{figure*}

\textcolor{black}{In Fig.~\ref{fig:BS_tTLG} the normal state band structure shown is shown in orange for reference. The band structure of tTLG is analogous to tBLG, but where there is an additional set of Dirac cones with large Fermi velocities to the flat bands, which has a magic-angle $\sqrt{2}$ times larger. As we are studying tTLG at charge neutrality, Hartree interactions can be neglected.}

The band structure of the FM state \textcolor{black}{of magic angle tTLG} is shown in Fig.~\ref{fig:BS_tTLG}(a). We found that this order can be stabilized with a Hubbard parameter of $U=1$ meV \textcolor{black}{at charge neutrality}. The pair of valence and conduction flat bands in the normal state are each split into a pair of spin-polarized bands, giving rise to four non-degenerate flat bands in this magnetic order, similar to the observations in tBLG. The splitting between the spin-polarized valence flat bands, in other words the bottom and second-highest band, is around 1 meV at the $\Gamma$-point and around 2 meV at the M-point. Such splitting is suppressed at the K(K')-points. However, there is a spin-degenerate band gap of around 3.1 meV due to time-reversal symmetry breaking in this order. 

By raising the Hubbard parameter to $U=2$ meV, the MMAFM order can also be stabilized, which is shown in Fig.~\ref{fig:BS_tTLG}(b). There is spin-degeneracy of bands, while a band gap at the K-point of around 0.93 meV persists. It does not, however, change the size of the band gaps at the $\Gamma$- and M-points. We found that these gaps further open for $U > 3$ meV. 

It is worth noting that, for both the FM and MMAFM orders, the graphene-like Dirac cones, i.e., those with a large Fermi velocity, remain unperturbed and do not hybridize with the flat bands. In general, for the FM and MMAFM orders, which can be stabilized with a relatively small $U$, the magnitude of the order parameters are around 0.1 meV. For the FM order, the constant background ($\delta_0$) in all layers is approximately twice the strength of the moir\'e-scale ($\delta_1$) oscillations. However, in the MMAFM order, the strengths of the constant ($\delta_0$) and moir\'e ($\delta_1$) part are almost equal in each of the layers. It is also found that the inner layer exhibits a stronger polarization than the outer layers, possibly because of the doubled moir\'e superlattice formed with the two outer layers, which enhances the moir\'e electronic effects specifically for the inner layer. 

\section{Discussion and Conclusions}

The self-consistent band structures we have obtained are consistent with those from Jimeno-Pozi \textit{et al}~\cite{Alejandro23} and Klebl \textit{et al}~\cite{Klebl_mag_order_tBLG_2019,Goodwin_mag_tBLG_2021}, who performed non-self-consistent calculations of tBLG. For example, the antiferromagnetic orders which break sublattice symmetry of the graphene layers cause the \textcolor{black}{a gap to open in the} Dirac cones, creating an insulating state at charge neutrality. \textcolor{black}{Previously, however, the ferromagnatic order was not investigated. } The ferromagnetic order does not break the sublattice symmetry, but it does break the spin degeneracy, causing the bands to spin-split, creating an insulating state at charge neutrality for tBLG. Overall, the self-consistent electronic structures obtained here qualitatively agree with those from  Jimeno-Pozi \textit{et al.}~\cite{Alejandro23} and Klebl \textit{et al}~\cite{Klebl_mag_order_tBLG_2019,Goodwin_mag_tBLG_2021}, and moreover, we have extended them to ferromagnetic order and tTLG.

By inspecting the regions of stability of the three competing ground state magnetic orders, there are also a number of notable similarities to the RPA predictions performed by Klebl \textit{et al.} in Ref.~\citenum{Goodwin_mag_tBLG_2021}. According to the RPA calculations carried out in Ref.~\citenum{Klebl_mag_order_tBLG_2019}, the critical values of Hubbard parameter $U_C$ reach a minimum near $\theta=1.08\degree..1.05 \degree$ both with Hartree and without Hartree interactions. This indicates that tBLG is susceptible to magnetic symmetry broken phases near the magic angle. Our predicted region of stability for MAFM and NAFM is consistent with this picture, in that the largest order parameters are found near the magic angle and decay away from it. Furthermore, in the RPA calculations it was found that FM order tends to occur at the critical $U_c$ for doped systems instead of M(N)AFM order, \textcolor{black}{which becomes strong with Hartree interactions}. Using the continuum model in this work we also find that the M(N)AFM order parameters are suppressed for larger doping levels\textcolor{black}{, and FM order can persist. Therefore, overall the self-consistent diagrams obtained here are qualitatively consistent with those by Klebl \textit{et al.}~\cite{Goodwin_mag_tBLG_2021}.}

\textcolor{black}{In the context of experimental measurements, our calculations reproduce some of the observed states, but not all of them. For tBLG, we predict a correlated insulating state at charge neutrality from anti-ferromagnetic order, owing to the sublattice symmetry breaking lifting the Dirac point, and also potentially an insulating state from ferromagnetic order. In experiments, we are not aware of any reports of a ferromagnetic state at charge neutrality, but the correlated insulating states observed at charge neutrality could have antiferromagnetic order~\cite{Lu2019,Cao2020,Cao2021,Das2021,Wu2021}. In the context of tTLG, the conclusion is similar, which agrees with experiments at charge neutrality~\cite{Hao21}. While our model has the pinning of van Hove singularities from Hartree interactions upon doping, we do not predict any other insulating states at other doping levels, for both tBLG~\cite{Jiang2019,Kerelsky2019,Xie2019} and tTLG. In contrast, experiments of tBLG and tTLG, which have correlated insulating states observed at practically all integer doping levels in the moir\'e bands~\cite{Lu2019}. Therefore, agreement with experiments is partial, which motivates further development of our approach. }

As the assumed magnetic orderings here do not break valley symmetries, creating insulating states at other doping levels than charge neutrality is not possible with the interactions included here. To go beyond this restriction, we must include longer-ranged exchange (Fock, for example) interactions, or modify the expression for the magnetic order parameters. For example, in tBLG we have constrained there to be a sum of cosines with certain polarisation on each sublattice, sometimes also including a sublattice polarised constant contribution. Instead, we could allow, for example, the sublattice polarisation to change in the constant and cosine contributions, allowing for more complex magnetic orderings. In the trilayer, many more instabilities were observed from the RPA calculations, and these could be included, or again more general forms of the magnetic instabilities included and self-consistently solved to find a ground state.

Overall, we have presented a formalism for the self-consistent inclusion of atomic-scale Hubbard interactions into the continuum model of moir\'e graphene multilayers. We have investigated this for tBLG and tTLG, for a number of magnetic instabilities, twist angles and doping levels. These developments can capture the interplay between long-ranged and short-ranged exchange interactions also in other moir\'e graphene multilayers, such as twisted double bilayer graphene. 

\section{Acknowledgments}

\textcolor{black}{We thank A Jimeno-Pozo and F Guinea for helpful discussions.} C. T. S. C. acknowledges funding from the Croucher Foundation and an Imperial College President's Scholarship. Z.A.H.G acknowledges support through the Glasstone Research Fellowship in Materials, University of Oxford. We acknowledge the Thomas Young Centre under Grant No. TYC-101. AF and DMK acknowledge funding by the DFG  within the Priority Program SPP 2244 ``2DMP'' -- 443274199. V. V. acknowledges support from the European Union NextGenerationEU - M4, C2, Investment line 1.2 - title "Ultrafast exciton dynamics and optoelectronics in moire superlattices - UltraDYNOMOS", CUP N. B53C24010860005. This  project  has  received  funding  from  the  European  Union’s  Horizon  2020  research  and  innovation programme under the Marie Sk\l{}odowska-Curie grant agreement No. 101067977. 

\section{Data and Code Availability}

The data which accompanies this work will be supplied upon request. The code is freely available under a GPL-3.0 license at \url{https://github.com/VVitale/TBLG-U}.

\bibliographystyle{apsrev4-1}
\bibliography{REF}
\end{document}